\documentclass[modern]{aastex61}

\newcommand\aastex{AAS\TeX}

\usepackage{graphicx}
\usepackage{epstopdf}

\accepted{10 December 2017}
\submitjournal{AJ}

\shorttitle{\aastex\ variability characterization of PSO J318.5-22}
\shortauthors{Biller et al.}
\begin{document}

\title{Simultaneous, multi-wavelength variability characterization of the free-floating planetary-mass object PSO J318.5-22}

\correspondingauthor{Beth Biller}
\email{bb@roe.ac.uk}

\author[0000-0003-4614-7035]{Beth A. Biller}
\affiliation{SUPA, Institute for Astronomy, University of Edinburgh, Blackford Hill, Edinburgh EH9 3HJ, UK}
\affiliation{Centre for Exoplanet Science, University of Edinburgh, Edinburgh, UK}

\author{Johanna Vos}
\affiliation{SUPA, Institute for Astronomy, University of Edinburgh, Blackford Hill, Edinburgh EH9 3HJ, UK}
\affiliation{Centre for Exoplanet Science, University of Edinburgh, Edinburgh, UK}

\author{Esther Buenzli}
\affiliation{Institute for Particle Physics and Astrophysics, ETH Zurich, Wolfgang-Pauli-Strasse 27, 8093 Zurich, Switzerland}

\author{Katelyn Allers}
\affiliation{Bucknell University; Department of Physics and Astronomy; Lewisburg, PA 17837, USA}

\author{Micka\"el Bonnefoy}
\affiliation{Institut de Plan\'etologie et d'Astrophysique de Grenoble, Universit\'e Grenoble Alpes, CS 40700, 38058 Grenoble C\'edex 9, France}

\author{Benjamin Charnay}
\affiliation{LESIA, Observatoire de Paris, PSL Research University, CNRS, Sorbonne Universit\'es, UPMC Univ. Paris 6, Universit\'e Paris-Diderot, Sorbonne Paris Cit\'e, 5 place Jules Janssen, 92195 Meudon, France} 

\author{Bruno B\'ezard}
\affiliation{LESIA, Observatoire de Paris, PSL Research University, CNRS, Sorbonne Universit\'es, UPMC Univ. Paris 6, Universit\'e Paris-Diderot, Sorbonne Paris Cit\'e, 5 place Jules Janssen, 92195 Meudon, France} 

\author{France Allard}
\affiliation{Univ Lyon, ENS de Lyon, Univ Lyon 1, CNRS, Centre de Recherche Astrophysique de Lyon UMR5574, F-69007 Lyon, France}

\author{Derek Homeier}
\affiliation{Zentrum f\"ur Astronomie der Universit\"at Heidelberg, Landessternwarte, K\"onigstuhl 12, D-69117 Heidelberg, Germany}

\author{Mariangela Bonavita}
\affiliation{SUPA, Institute for Astronomy, University of Edinburgh, Blackford Hill, Edinburgh EH9 3HJ, UK}
\affiliation{Centre for Exoplanet Science, University of Edinburgh, Edinburgh, UK}

\author{Wolfgang Brandner}
\affiliation{Max-Planck-Institut f\"ur Astronomie, K"onigstuhl 17, D-69117 Heidelberg, Germany}

\author{Ian Crossfield}
\affiliation{Department of Physics, Massachusetts Institute of Technology, 77 Massachusetts Avenue, Cambridge, MA 02139, USA}

\author{Trent Dupuy}
\affiliation{Gemini Observatory, 670 N. Aohoku Pl., Hilo, HI 96720, USA}

\author{Thomas Henning}
\affiliation{Max-Planck-Institut f\"ur Astronomie, K"onigstuhl 17, D-69117 Heidelberg, Germany}

\author{Taisiya Kopytova}
\affiliation{School of Earth \& Space Exploration, Arizona State University, Tempe AZ 85287, USA}

\author{Michael C. Liu}
\affiliation{Institute for Astronomy, University of Hawaii at Manoa, 2680 Woodlawn Drive, Honolulu, HI, 96822, USA}

\author{Elena Manjavacas}
\affiliation{Steward Observatory, University of Arizona, 933 N. Cherry Ave, Tucson, AZ 85917, USA}

\author{Joshua Schlieder}
\affiliation{Exoplanets and Stellar Astrophysics Laboratory, Code 667, NASA Goddard Space Flight Center, Greenbelt, MD, USA}



\begin{abstract}
%Variability attributed to cloud structure is a persistent feature for L and T type field brown dwarfs. Directly imaged planets occupy the same temperature regime as L and T type brown dwarfs and are likely to be equally variable, although the characteristics of their variability may differ from that of field brown dwarfs due to the low surface gravity of these objects.  While few exoplanet companions are amenable to variability searches, a wider class of "exoplanet analogues" can be defined with similar spectral types and surface gravities as young exoplanet companions.  
We present simultaneous HST WFC3 + Spitzer IRAC variability monitoring for the highly-variable young ($\sim$20 Myr) planetary-mass object PSO J318.5-22.  Our simultaneous HST + Spitzer observations covered $\sim$2 rotation periods with Spitzer and most of a rotation period with HST.  We derive a period of 8.6$\pm$0.1 hours from the Spitzer lightcurve.  Combining this period with the measured $v sin i$ for this object, we find an inclination of 56.2$\pm 8.1^{\circ}$.  We measure peak-to-trough variability amplitudes of 3.4$\pm$0.1$\%$ for Spitzer Channel 2 and 4.4 - 5.8$\%$ (typical 68$\%$ confidence errors of $\sim$0.3$\%$) in the near-IR bands (1.07-1.67 $\mu$m) covered by the WFC3 G141 prism -- the mid-IR variability amplitude for PSO J318.5-22 one of the highest variability amplitudes measured in the mid-IR for any brown dwarf or planetary mass object.  Additionally, we detect phase offsets ranging from 200--210$^{\circ}$ (typical error of $\sim$4$^{\circ}$) between synthesized near-IR lightcurves and the Spitzer mid-IR lightcurve, likely indicating depth-dependent longitudinal atmospheric structure in this atmosphere.  The detection of similar variability amplitudes in wide spectral bands relative to absorption features suggests that the driver of the variability may be inhomogeneous clouds (perhaps a patchy haze layer over thick clouds), as opposed to hot spots or compositional inhomogeneities at the top-of-atmosphere level.  
\end{abstract}




\section{Introduction} \label{sec:intro}

Rotationally-modulated variability is a key probe of exoplanet and brown dwarf atmospheric structure.  Field brown dwarfs are generally rapid rotators, with periods of 3-20 hours \citep{Zap06}.  Several young planetary mass objects now have measured rotation periods $<$20 hours as well \citep[][]{Sne14, Bil15,Lew16,Zho16,Vos17b}.
Any top-of-atmosphere inhomogeneity on rapid rotators (due e.g. to cloud structure, thermo-chemical instabilities, or auroral emission) may be detectable via the quasiperiodic photometric variability it produces.  Variability is common in high surface gravity field brown dwarfs, with the maximum variability amplitude in the near-IR at the L/T spectral type transition 
\citep{Rad14a,Rad14b,Wil14,Met15}. Most efforts to model the mechanism driving the observed quasiperiodic variability have utilized patchy thin and thick cloud cover \citep{Apa13}, possibly related to the dissipation of clouds across the L/T transition.  Variability studies enable a multi-dimensional view of these atmospheres -- lightcurves at shorter wavelengths generally probe deeper atmospheric pressure levels compared to longer wavelengths \citep{Bue12,Bil13,Yan16} and the rapid rotation of these objects move different parts of their atmospheres in and out of view \citep{Apa13}.    

Due to the extreme contrast difference between host star and planet, very few exoplanet companions are amenable to high-precision variability searches.  However, a wider class of "exoplanet analogues" can be defined with similar spectral types and surface gravities
as young exoplanet companions. Several young planets and exoplanet analogues also have measured periods $\leq$20 hours \citep{Sne14,Bil15,All16,Zho16}, allowing variability searches with reasonable observation lengths from both ground and space.
Known exoplanet analogues include bonafide free-floating planetary mass objects such as the 8.3$\pm$0.5 M$_{Jup}$~VL-G L7$\pm$1~$\beta$ Pic moving group member PSO J318.5-22 \citep{Liu13,All16} as well as slightly higher mass ($<$25 M$_{Jup}$), young objects such as WISEP J004701.06+680352.1 (henceforth W0047), a $\sim$20 M$_{Jup}$, very red AB Dor (150 Myr) moving group member \citep[][]{Giz12,Giz15,Fili15,Vos17b}. While they likely do not share a formation mechanism with exoplanet companions, they share similar masses and surface gravities \citep{Fah16}.  Thus, they allow us to study similar atmospheres without first overcoming the light of a nearby star.  In particular, PSO J318.5-22 and W0047 have spectra that are nearly identical to the inner two HR 8799 planets \citep{Bon16}.  The low surface gravity in
young objects significantly affects the spectra of these objects and also the T$_\mathrm{eff}$ at which these objects would transition between the L and T spectral types.  Low surface gravity objects have redder colors compared to field dwarfs and potentially retain thick silicate clouds or small-scale turbulent energy transport via thermo-chemical instabilities down to lower T$_\mathrm{eff}$ than field counterparts with similar spectral types \citep{Bar11,Liu16,Tre16}.

Variability has recently been detected for the first time in young planetary mass objects. 
Vos et al. (in prep) surveyed $\sim$40 low surface gravity planetary mass objects and very young brown dwarfs with L and T spectral types with NTT SofI and UKIRT WFCAM (sensitive to variability amplitudes $>$2$\%$), detecting low-amplitude variability in 3 L1-L4 objects and high-amplitude variability (7-10$\%$) in the mid-to-late L, low surface gravity objects PSO J318.5-22 \citep{Bil15} and 2MASS J2244316+204343 \citep[henceforth~2M2244,][]{Vos17b}, a $\sim$20 M$_{Jup}$, AB Dor (150 Myr) moving group member \citep[][]{Fili15,Vos17b}.  \citet{Zho16} report lower-amplitude variability in the mid-L companion planetary mass object 2MASSW J1207334-393254b (henceforth 2MASS 1207b) and  \citet{Lew16} found high amplitude variability (8$\%$) for WISE 0047.  For young, T-spectral type objects, two detections have been reported to date. Recently, \citet{Gagn17} identified the highly variable T2.5 dwarf SIMP 0136 \citep{Art09} as a likely member of the 200 Myr Carina-Near moving group.  \citet{Naud17} report a tentative J-band detection of variability for the wide T3.5 planetary mass companion GU Psc b, with 4$\pm$1$\%$ variability on a six-hour timescale detected in one epoch out of three of monitoring.  

The mid-to-late M, low-surface gravity objects PSO J318.5-22, WISE 0047, and 2M2244 have the highest near-IR amplitudes ($>$5$\%$) measured for {\it any} L type object to date.  %In contrast, the detection of variability in the L5 planetary mass object 2M1207b has a lower amplitude of 1-2$\%$ 
%(Zhou et al. 2016).  
Until recently few late L low surface gravity objects had been identified and only four such objects with spectral types between L6.5 and L9 have been monitored for variability in the near-IR \citep[][Vos~et~al.~in~prep]{Mor06,Lew16,Bil15,Vos17b}.
%Thus, it is not clear whether PSO J318.5-22 and W0047 are unusual -- or if the peak of near-IR variability for young, low surface gravity objects is at late-L spectral types (instead of L/T transition spectral types for field dwarfs).  
From model predictions, these late-L objects are expected to have thick (and probably homogeneous) cloud cover \citep[e.g.][]{Mad11}, although some recent modeling efforts have posited that the red colors of late-L dwarfs could alternatively be produced as the result of convection driven by thermo-chemical instabilities \citep{Tre16,Trem17}.  Late-L low surface gravity objects already clearly demonstrate differences in variability properties compared to field brown dwarfs, which peak in variability amplitude at later spectral types (T0-T2) along the L/T spectral type transition \citep{Rad14a,Rad14b}, potentially due to breakup or patchiness of clouds at this spectral type transition. 
These differences are likely a result of the lower surface gravities for the young, planetary mass late-L objects compared to field brown dwarfs.  Studying the variability of young objects in detail will illuminate the role of surface gravity in determining atmospheric structure.

%The current cohort of imaged giant exoplanets as well as higher mass brown dwarfs fill out a grid of atmospheres with fundamental variables of mass, effective temperature, composition, and surface gravity.  Comparative studies of different atmospheres along this grid help us understand the properties of these atmospheres as a whole.

Simultaneous multi-wavelength variability monitoring is a powerful
tool to understand the atmospheres of these objects, 
allowing us to pinpoint the mechanism driving the variability.
Different wavelength regimes probe different atmospheric depths \citep{Mar12} and high spectroscopic resolution allows us to study
variability within individual spectral features.  
\citet{Mor14b} find that variability due to patchy
clouds should drive high amplitude variability within wide spectral
windows while variability due to hot spots (i.e. heating at a specific
pressure level) should drive larger variability within absorption
features.  Based on this, \citet{Mor14b} suggest that simultaneous, multi-wavelength observations 
probing both inside and outside molecular absorption features will prove to be particularly valuable in understanding the physical processes driving this variability.
%``the most illustrative types of observations for understanding the physical processes
%underlying brown dwarf variability are simultaneous, multi-wavelength observations that probe both inside and%
%outside of molecular absorption features.'' 

Only $\sim$10 old field dwarfs have published HST spectroscopic variability
monitoring \citep{Bue15a,Bue15b,Apa13,Bue12,Yan15,Yan16}. SIMP 0136 \citep{Apa13} and W0047 \citep{Lew16} are the only potentially young objects with published HST spectroscopic variability monitoring.  HST enables exceptionally high photometric precision as well as access
to the 1.4 and 1.1 $\mu$m water absorption bands, which are
unobservable from the ground due to telluric absorption. 
For L/T transition objects such as 
Luhman 16B, SIMP 0136, and 2M 2139 \citep{Bue15a,Apa13}, these studies have found
correlated variability across the J and H band, with decreased
variability in the 1.4 $\mu$m water absorption feature -- 
consistent with variability due to inhomogeneous thick and thin cloud
cover. For the T6.5 2M 2228, \citet{Bue12} found significant
phase shifts at different wavelengths, including between broadband J
and the 1.4 $\mu$m water feature -- interpreted as differences in
cloud properties at different atmospheric levels.  For the mid-L
dwarfs, 2M 1821 and 2M 1507, \citet{Yan15} found correlated
variability with similar amplitude in the 1.4 $\mu$m water band as in 
broadband J and H -- this is interpreted as variability due to high
level hazes (above significant water concentrations) in these atmospheres. 

Here we present simultaneous HST WFC3 + Spitzer IRAC variability monitoring 
for the variable planetary-mass object PSO J318.5-22.
Of the current ensemble of free-floating young objects 
with estimated masses $<$30 M$_{Jup}$ , PSO J318.5-22 \citep{Liu13} is the closest analogue in properties to imaged
exoplanet companions.   \citet{Gag14} and \citet{Liu13} identified it as a
$\beta$ Pic moving group member and it possesses 
colors and magnitudes similar to the HR 8799 planets and 2M1207b \citep{Bon16}.
Using evolutionary models and adopting an age of 23$\pm$3 Myr,
\citet{All16} find an effective temperature of 
T$_{eff}$ = 1127$^{+24}_{-26}$ K and a mass estimate of
8.3$\pm$0.5 M$_{Jup}$ for PSO J318.5-22.  Understanding the variability of 
this benchmark object will yield fundamental insights into its atmospheric
properties, especially regarding the presence of clouds -- and by
proxy the expected properties of exoplanet companion atmospheres.

\section{Observations}

Simultaneous HST + Spitzer observations of PSO J318.5-22 were acquired on 8-9 September 2016.
Spitzer observations lasted from UTC 2016-09-08 09:01:14 to UTC 2016-09-09 02:18:27, with 5 HST orbits taken from UTC 2016-09-08 11:38:59 to UTC 2016-09-08 18:44:41, for a simultaneous monitoring period of $\sim$7 hours, and a Spitzer monitoring period of $\sim$17.2 hours.  

Spitzer observations were taken with IRAC in Channel 2 (4.5 $\mu$m) in staring mode, with 1940$
\times$30 s frames acquired (program id 12002).  As Spitzer requires $\sim$30 minutes to settle after a target is acquired, a short dithered sky sequence (9$\times$30 s frames taken at 5 dither positions) preceded the science sequence.  A short sky sequence (1$\times$30 s frame taken at 5 dither positions) was acquired after the science sequence as well.  Following established procedures to ensure optimal photometric precision and correct for intrapixel sensitivity variations \citep{Migh08}, care was taken to place the target in the IRAC "sweet spot" during the science sequence, which lies in the upper left quadrant of the full detector.  

HST observations were taken with the infrared channel of WFC3 with the G141 grism (program ID 14188).  The full 123$\times$136 arcsec frame was used with the SPARS25 readout mode, enabling observation of 6 background stars as well as the target.  Each 90-minute orbit yielded 59 minutes of usable exposure time, when the target was not occulted by the Earth.   To determine object positions on the detector for each orbit, a direct image was taken in the F127M filter with exposure time of 53 s (NSAMP=3).  Thereafter, a sequence of 9$\times$278 s (NSAMP=12) exposures were taken with the G141 grism, which covers a wavelength range of 1.077 $\mu$m to 1.7 $\mu$m, with resolution R=130 at 1.4 $\mu$m.  With the remaining orbital visibility, an additional 53~s (NSAMP=3) grism exposure was taken at the end of orbit 1 and an additional 153 s grism exposure (NSAMP=7) was taken in orbits 2-5.  These final, shorter exposures were significantly noisier than the other exposures and were omitted from the final analysis.  
One 278 s exposure in orbit 4 suffered complete data loss, as the data were not fully read off the HST recorder before being overwritten, and is thus omitted in the following analysis. 

\section{Spitzer data reduction and lightcurve extraction}

No Spitzer [4.5 $\mu$] magnitude has previously been reported in the literature for PSO J318.5-22.  From the full-sequence Spitzer MOSAIC images, we derived a Spitzer [4.5 $\mu$] magnitude of 12.541$\pm$0.017, using the code described in \citet{Dup13}.  This is in good agreement with the WISE W2 magnitude for this object \citep{Liu13}.

To construct a lightcurve as a function of time, we extracted photometry from the corrected Basic Calibrated Data images from the Spitzer Science Center, 
processed with IRAC pipeline version 19.2.0. 
Times for each photometric point were taken from the $MJD\_OBS$ header keyword, which provides the Modified Julian Date.
After finding centroids for the target and a number of reference stars
using box$\_$centroider.pro, we performed aperture photometry 
about these centroids.  A range of apertures were tested -- 
we adopt here an aperture of 2.4 pixels, which produced lightcurves with 
the lowest RMS.

To robustly remove outliers while avoiding 
subtracting out any astrophysical variability, 
we followed the clipping procedure described in detail in 
Heinze et al. (2014).  We median-smoothed each lightcurve
with a sliding boxcar (width of 25 frames, corresponding to 12.5 minutes).
The smoothed lightcurve is subtracted from the original data, 
removing astrophysical and systematic signals with timescales 
longer than 12 minutes.  Thus, any outliers remaining 
in the subtracted lightcurve must be artifacts and can be 
confidently removed, using a 6-$\sigma$ clip.

The flux of an object on a given point of the IRAC detector will vary 
depending on exactly where a point source falls with respect to the 
centre of a pixel -- this is known as the 'pixel phase effect'.  
We correct for the pixel phase effect using the pixel$\_$phase$\_$correct$\_$gauss.pro
routine from the Spitzer IRAC website, which models the pixel phase response 
as a double-gaussian, a summation of gaussians in the orthogonal pixel directions.
The pixel phase corrected 
flux is then binned into 2.5 min bins.  

\section{HST data reduction and lightcurve extraction}

We extracted spectra for PSO J318.5-22 and 6 background stars on the detector from FLT calibrated individual exposures downloaded from the MAST archive.  Times for each exposure were taken from the mean of the $EXPSTART$ and $EXPEND$ header keywords, which are provided as Modified Julian Date.  The FLT files have been processed using $calwfc3$, which performs basic data calibration including bad pixel flagging.  We then corrected for bad pixels flagged by $calwfc3$, specifically pixels flagged with flag values 1, 4, 32, and 512.  Bad pixel correction was performed by interpolating over the pixels on the left and the right of the flagged pixel.  If the right-side pixel was also flagged, then only the left pixel is used to correct the original flagged pixel.  We visually identified one bad pixel in the spectrum of PSO J318.5-22 that was not automatically flagged by $calwfc3$.  This pixel was corrected manually. 

For the direct images taken at the beginning of each orbit, object positions on the chip were obtained using Source Extractor (SExtractor).  For each object on the detector, the aXe pipeline was then used to extract slitless spectroscopy from each of the 9$\times$278 s grism exposures per orbit, using the FLT grism files and the object positions obtained with SExtractor as inputs.  Sky subtraction was performed using the $aXeprep$ routine and spectrum extraction was performed using the $aXecore$ routine.  The usable spectral bandwidth runs from 1.07 to 1.67 $\mu$m, with a resolution of R=130.  We extracted spectra with extraction widths ranging from 1 - 20 pixels.  We found that a 7 pixel extraction width best balanced object signal against background noise.  Thus, we adopt the 7-pixel extraction width for subsequent analysis.  We flux-calibrated the extracted spectra using the G141 sensitivity curve.  
%(Time -- is it MJD or barycentric MJD, look this up...)  

We extracted lightcurves from the spectra across a variety of spectral bandwidths.  We integrated over the full 1.07-1.67 $\mu$m spectral range to generate a "white-light" lightcurve.  In order to compare with ground-based studies, we also integrated over the standard 2MASS J and H bandpasses.  Note that the 2MASS H band extends to wavelengths longer than the 1.67 $\mu$m cutoff for the G141 grism, thus we have only integrated over the portion of the 2MASS H filter that falls within the G141 grism spectral range.  We consider variability as well in two spectral features -- integrating from 1.34 to 1.44 $\mu$m to capture variability in the 1.4 $\mu$m water absorption feature and from 1.60 to 1.67 $\mu$m to capture variability in the 1.6 $\mu$m methane absorption feature.

As noted by previous studies \citep{Bue12, Apa13,Bue15a}, WFC3 photometry displays a "ramp effect" -- where the flux appears to increase with an exponential ramp at the beginning of each orbit.  This is especially notable in the first orbit of a visit.  \citet{Bue12} find this effect to be independent of count rate and wavelength.  Previous authors who have utilized a 256x256 pixel subarray instead of the 1048x1048 full frame have corrected this effect by using an analytic function derived from a non-variable source in the field \citep{Apa13}.  Because we have 6  background stars in the field,
we choose to build a correction based solely on these background stars, without fitting an analytic function or a detector-based model as described in \citet{Zhou17}.  Of the 6 background stars, one is considerably fainter than other objects on the detector and another appears to be somewhat variable itself.  For the remaining 4 well-behaved stars (all 2-3$\times$ as bright as the target object), we combine their normalized white light lightcurves to produce a calibration curve (utilizing median combination, then taking the average of the two central values, as we use an even number of reference stars).  We then divide both target and background star lightcurves and spectra by the calibration curve to correct for the ramp effect as well as other systematics which affect all objects on the detector \citep[this~is~similar~to~the~approach~taken~in~ground~based~studies~such~as][]{Rad14a, Bil15}.

%Other than PSO J318.5-22, there are 
%6 relatively bright background stars on the detector, which we eventually use to detrend the data.

\section{Results}

\begin{figure*}
% For example, with the graphicx style use
\includegraphics[width=\linewidth]{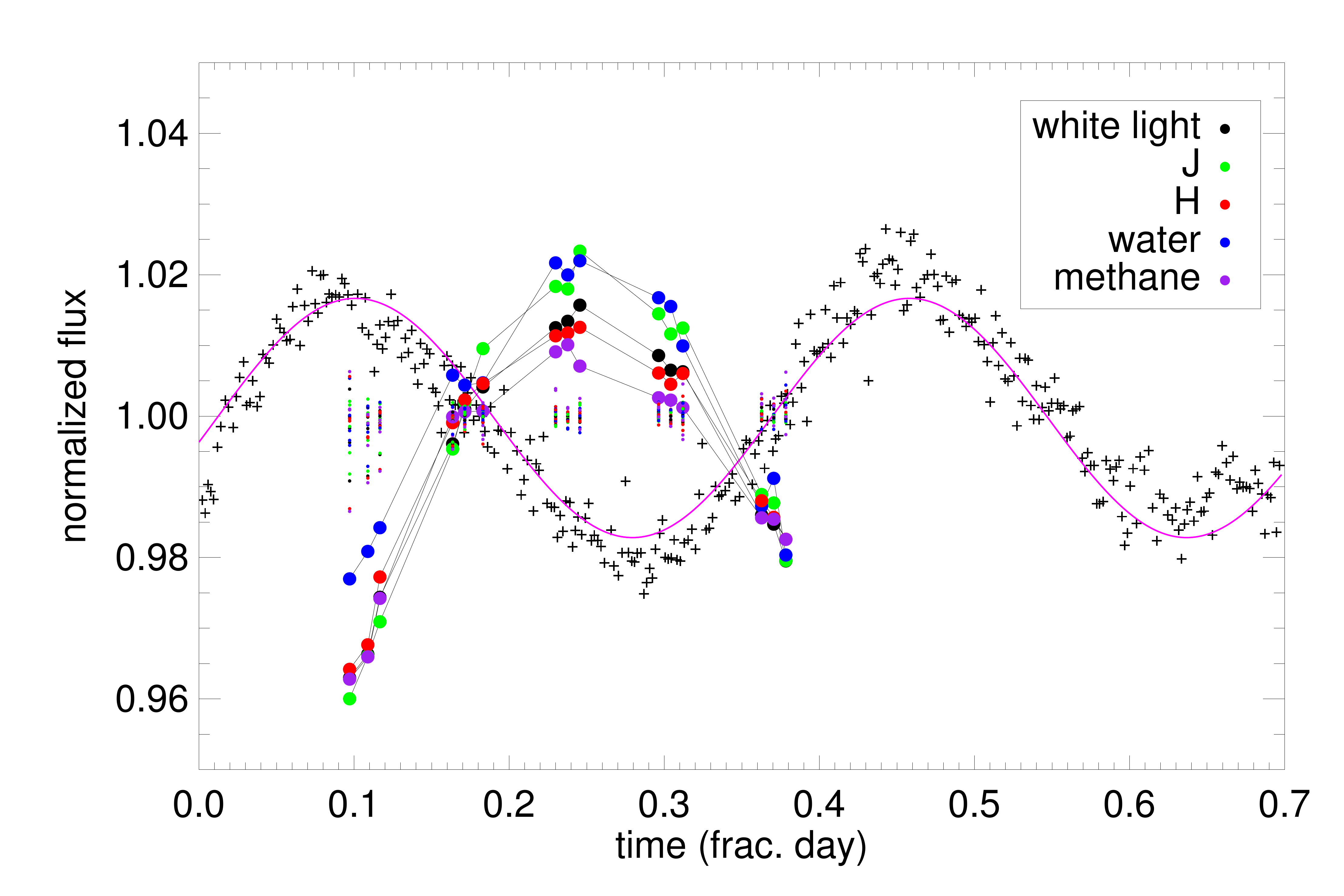}
\caption{Spitzer (crosses) and HST lightcurves (filled circles) for PSO J318.5-22, after correction for the ramp effect.  The lightcurves have been binned to increase S/N ratio, resulting in a 2.5 minute cadence for Spitzer and a 14 minute cadence for HST.  The least-squares best fit to the Spitzer lightcurve is shown as a solid purple line. HST lightcurves are shown binned over 5 spectral bandwidths: the full usable 1.07 - 1.67 $\mu$m spectral bandwidth of the HST grism spectroscopy (white light, black circles), the 2MASS J band (green circles), the 2MASS H band up to the spectral cutoff at 1.67 $\mu$m (red circles), a band centered on the 1.4 $\mu$m water absorption feature (blue circles), and a band covering as much of the 1.6 $\mu$m methane absorption features as falls in the HST G141 grism spectral bandwidth (purple circles).  Small colored points are the 6 background stars in the HST field after being detrended by the calibration curve; PSO J318.5-22 is clearly variable compared to the reference stars.  The large 200-210$^{\circ}$ phase offsets between the near-IR and mid-IR lightcurves likely indicates varying longitudinal atmospheric structure at different depths in this atmosphere.
}
\label{fig:1}       % Give a unique label
\end{figure*}

Spitzer and HST lightcurves (after correction for the ramp effect) for PSO J318.5-22 are presented in Fig.~\ref{fig:1}.  To increase the S/N ratio, the lightcurves have been binned by a factor of 5 for Spitzer, resulting in a 2.5-minute cadence, and by a factor of 3 for HST, resulting in a 14-minute cadence.  Small colored points are the 6 background stars after being detrended by the calibration curve; PSO J318.5-22 is clearly variable compared to the reference stars.

\begin{figure*}
% For example, with the graphicx style use
%\includegraphics[scale=0.9]{spectrum_comparison.pdf}
\includegraphics[width=0.9\linewidth]{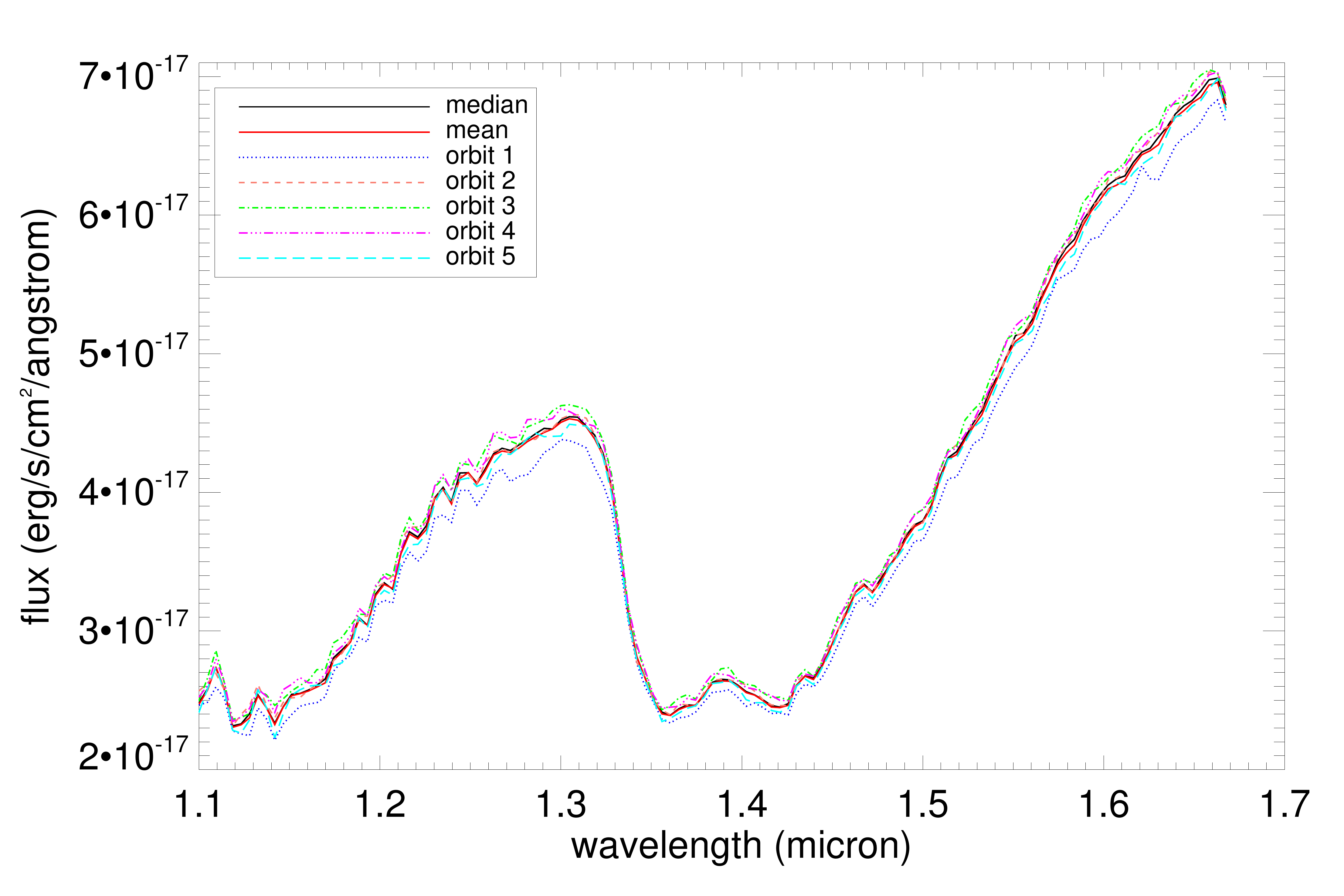}
\includegraphics[width=0.9\linewidth]{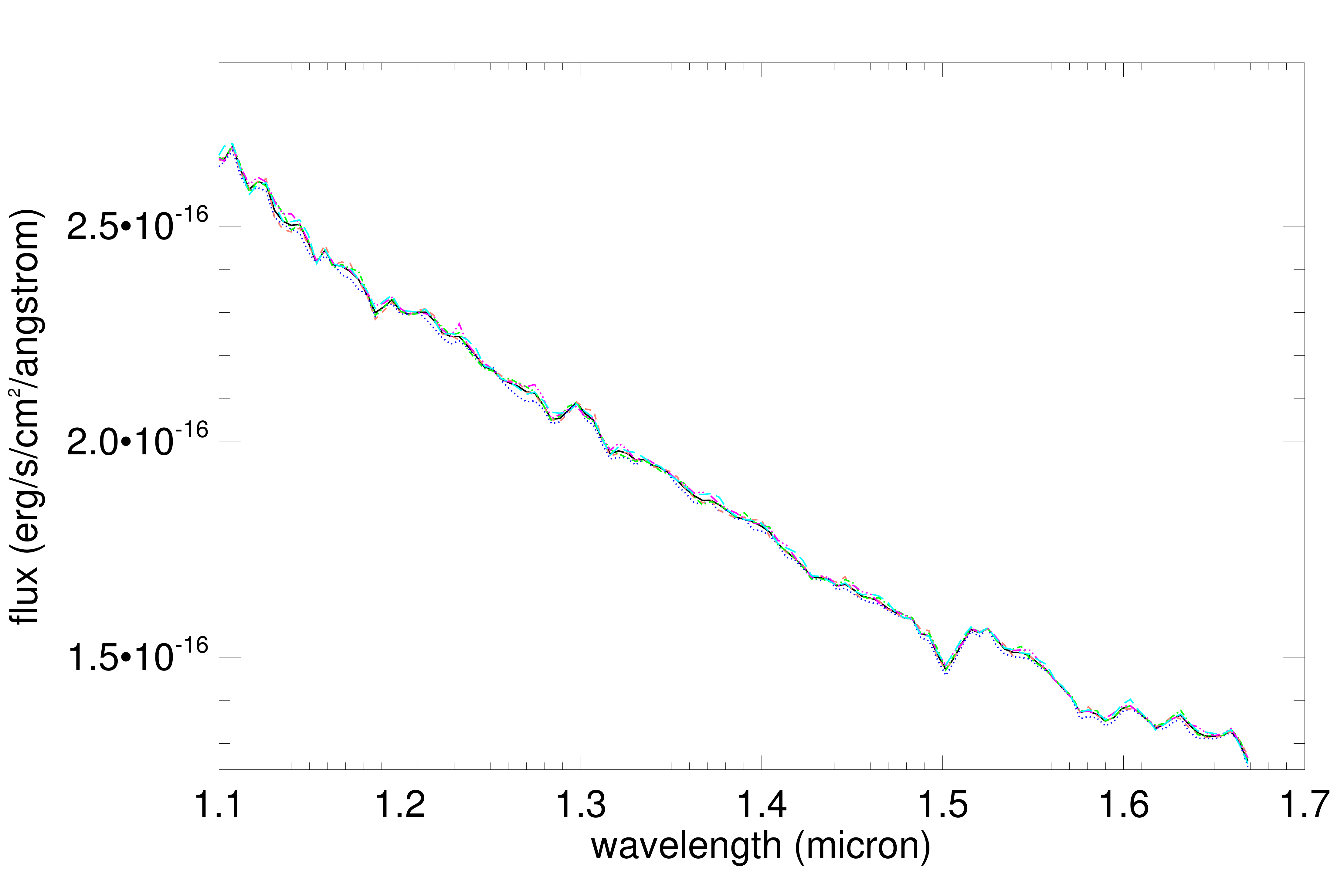}
\caption{{\bf Top:} The mean and median spectra across the full 5 orbit HST
observation. Significant spectral variability is apparent.  
{\bf Bottom:} Similar spectra for one of the well-behaved, non-variable
reference stars in the HST field.  The legend is shared between both panels.
}
\label{fig:2}       % Give a unique label
\end{figure*}

The mean and median spectra across the full 5 orbit HST observation, as well as the median spectrum per orbit, are presented in the top panel of Fig.~\ref{fig:2}.  Similar spectra for one of the well-behaved reference stars is shown in the bottom panel of Fig.~\ref{fig:2}.

\subsection{Period and amplitude from the Spitzer light curve}

\begin{figure*}
% For example, with the graphicx style use
\includegraphics[width=\linewidth]{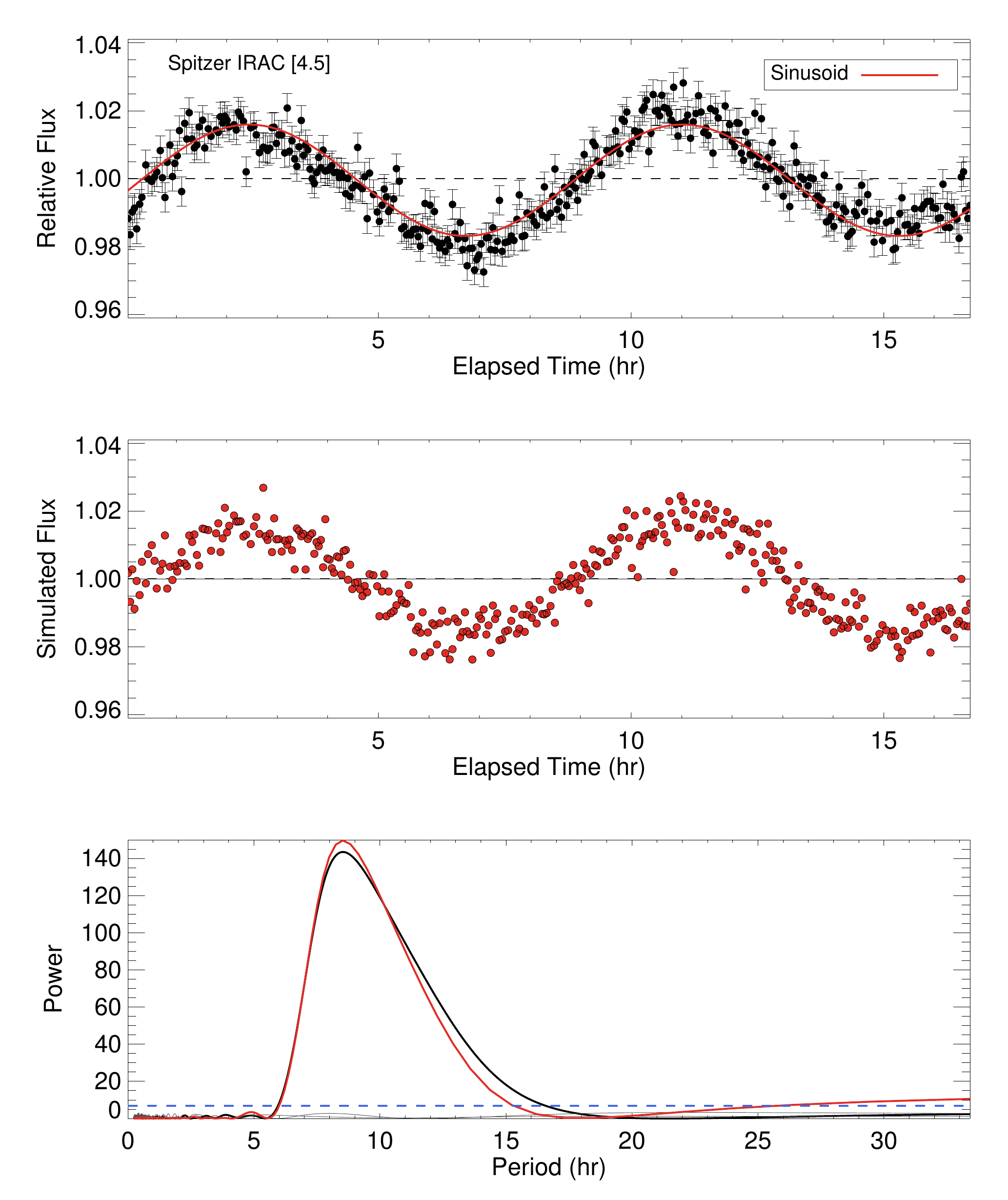}
\caption{The top panel shows the normalized, pixel phase corrected
lightcurve of PSO J318.5-22 with best-fit sinusoidal function overplotted
in red.  The middle panel shows the best fit
function with Gaussian noise added -- this simulated lightcurve closely resembles the observed lightcurve.  The bottom panel
shows the periodogram of the target and the simulated curve, as
well as the periodogram of several reference stars in the field. The
blue dashed line shows the 1\% false-alarm probability.
}
\label{fig:periodogram}       % Give a unique label
\end{figure*}

The unbinned Spitzer lightcurve (30 s cadence) along with the best fit sinusoid using a Levenberg-Marquardt least-squares minimization algorithm 
is presented in Fig.~\ref{fig:periodogram}.  The sinusoidal model has four parameters: period (in hours), phase (in degrees), mean lightcurve value (since we have divided the raw lightcurve by the median flux over the whole observation, this should tend towards unity), and amplitude (in percent variation, peak to mean lightcurve value).  The best fit model with gaussian noise added is also shown and provides a good match to the observed lightcurve.  We also plot the periodogram in Fig.~\ref{fig:periodogram} of PSO J318.5-22 as well as a number of reference stars in the field to identify periodic variability in our targets. The 1\% false-alarm probability (FAP, plotted in blue on the figure) is calculated from 1000 simulated lightcurves. These lightcurves are produced by randomly
permuting the indices of reference star lightcurves \citep{Rad14a}, producing lightcurves with Gaussian-distributed noise.  The observed variability is reasonably well modeled with a sinusoidal model, although successive maxima and minima appear to be marginally increasing in flux.  Thus we considered as well sinusoidal + linear models.

\begin{figure*}
% For example, with the graphicx style use
\includegraphics[scale=0.6]{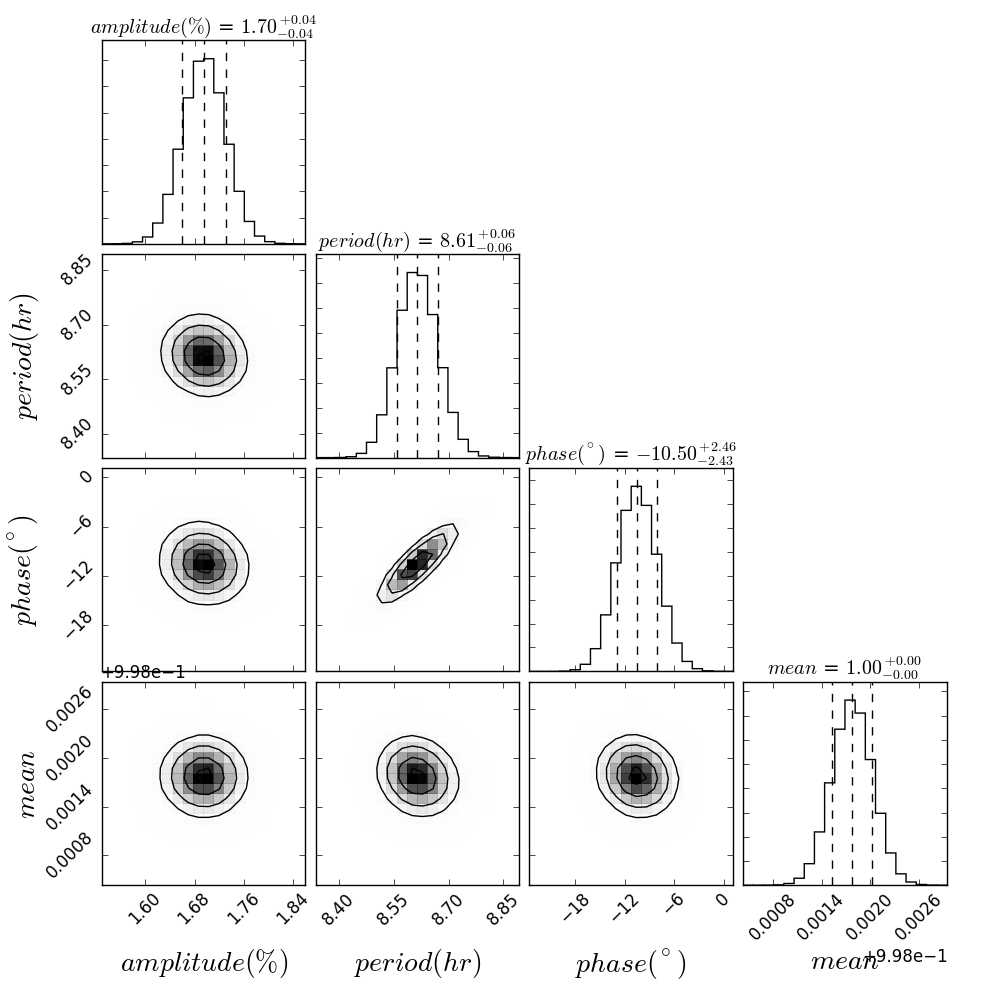}
\caption{Posterior probability distributions of parameters from sinusoid MCMC fits to our Spitzer Channel 2 lightcurve of PSO J318.5-22.  The mean parameter is the mean
value of the lightcurve -- since we have divided the raw lightcurve by the median flux over the whole observation, this should tend towards unity.  In the marginalized confidence interval plots, the middle dashed line gives the median, the two outer vertical dashed lines represent the 68\% confidence interval. The contours show the 1, 1.5 and 2-$\sigma$ levels.
}
\label{fig:sinusoid}       % Give a unique label
\end{figure*}

\begin{figure*}
% For example, with the graphicx style use
\includegraphics[scale=0.5]{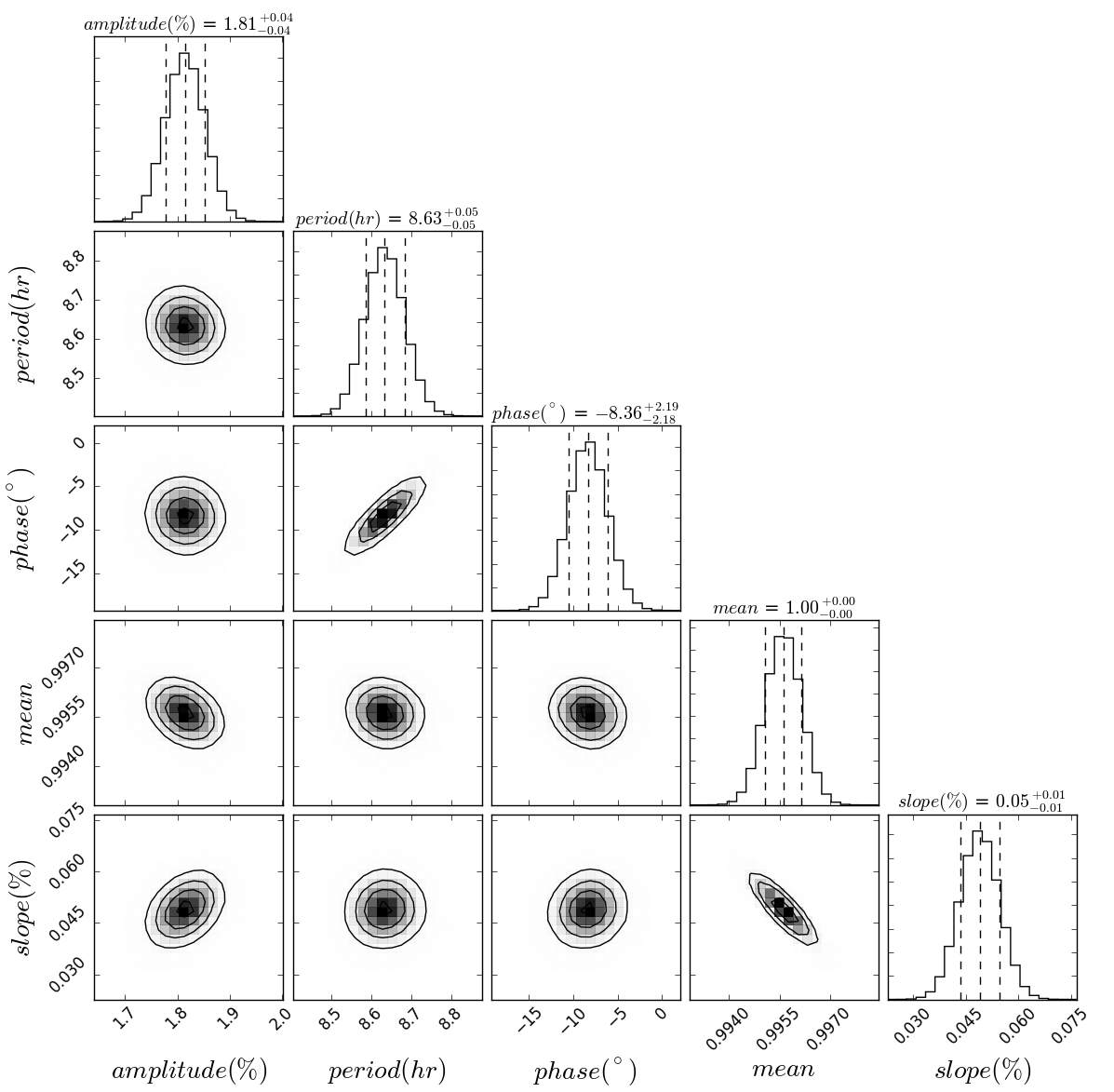}
\caption{Posterior probability distributions of parameters from sinusoid+linear MCMC fits to our Spitzer Channel 2 lightcurve of PSO J318.5-22.  In the marginalized confidence interval plots, the middle dashed line gives the median, the two outer vertical dashed lines represent the 68\% confidence interval. The contours show the 1, 1.5 and 2-$\sigma$ levels.
}
\label{fig:sinusoidslope}       % Give a unique label
\end{figure*}

\begin{figure*}
% For example, with the graphicx style use
\includegraphics[scale=0.8]{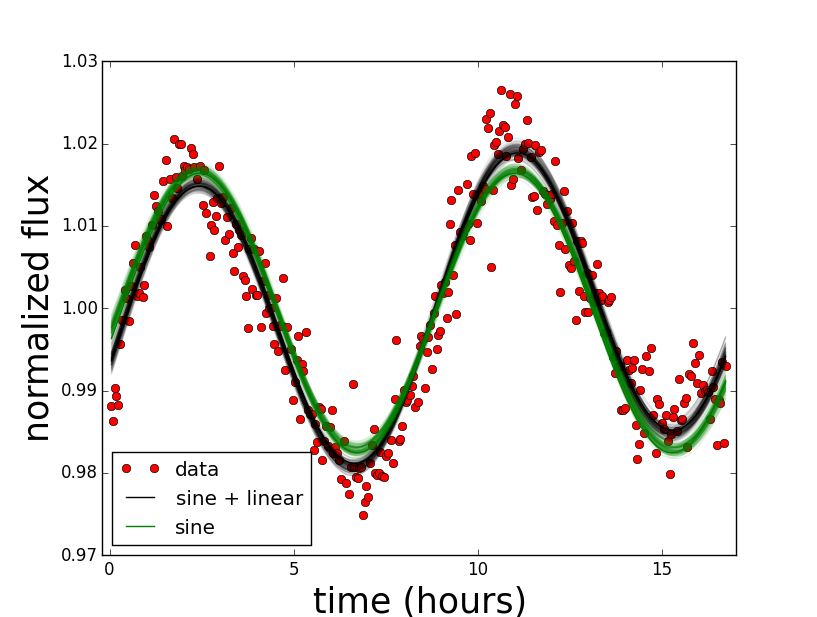}
\caption{100 sample model lightcurves drawn respectively from our sinusoid (green) and sinusoid+linear (black) MCMC fits to the Spitzer Channel 2 lightcurve (plotted as filled red circles).  
}
\label{fig:samples}       % Give a unique label
\end{figure*}

To fully explore the parameter space of both sinusoidal and sinusoidal + linear fits, we used the {\it emcee} Markov-Chain Monte Carlo package \citep{Fore13} to determine the full posterior probability distribution.  We ran an MCMC chain using a $\chi^2$ likelihood function with 1000 walkers for 2000 steps.  The first 100 steps of each chain were thrown out as part of the "burn-in".  Chains were checked by eye for convergence.  We fit both a single sinusoid model (results shown in Fig.~\ref{fig:sinusoid}) and a sinusoid + linear model (shown in Fig.~\ref{fig:sinusoidslope}).  We adopt the 50$\%$ quantile value as the best value for amplitude and period.  Best values of amplitude (peak-to-median value) and period as well as 68$\%$ confidence-interval and 95$\%$ confidence-interval errors are presented in Table~\ref{tab:Spitzermcmc}.  Both provide reasonably good fits -- in Fig.~\ref{fig:samples}, we plot 1000 samples from our chain for both models.  We calculated the Bayesian Information Criterion \citep[BIC,][]{schwarz1978} for the adopted best value parameters for both the single sinusoid and sinusoid + linear fit.  The BIC is given by:

\begin{equation}
BIC = ln(n)~k - 2~ln(\mathcal{L})
\end{equation}

where $n$ is the number of data points in the lightcurve (1700 for the Spitzer curve), $k$ is the number of model parameters, and $\mathcal{L}$ is the maximized value of the likelihood function of the model.  We find BIC$\simeq$12 for the single sinusoid and BIC$\simeq$19 for the sinusoid + linear fit.   The model with the lower value of BIC is preferred, thus we adopt the sinusoid-only model for further analysis and comparison to HST results.  

\citet{Apai17} recently modeled the rapid brightness evolution found in brown dwarf lightcurves using beating patterns between multiple planetary-scale wave surface features that move at slightly different velocities due to zonal wind speed variations.  They considered both a simple, 3-sinusoid model and applied as well their Aeolus mapping package, with both spots and planetary-scale waves \citep{Kar16}.  We attempted to fit a similar, 3-sinusoid model to our Spitzer lightcurves, with the periods of the three sinusoids given as: $P_{rot} = (P_1 + P_2) / 2$ (wavenumber k=1 waves) and $P_3 = P_{rot}/ 2$ (k=2 wave), where P$_{rot}$  is the rotational period of PSO J318.5-22.  In this case, the best fitting model always reverted to a single sinusoid, with negligible amplitudes for the other two sinusoids relative to the uncertainties in the lightcurve.  We only cover $\sim$2 apparent rotational periods for PSO J318.5-22, thus it is unclear from our Spitzer lightcurve alone whether the nearly sinusoidal variation observed is the fundamental lightcurve of this object or whether we have observed it during a period when multiple planetary-scale wave surface features happen to be in phase.  This is qualitatively similar to the case of SIMP 0136 described in \citet{Apai17}, where the apparent periods of the two k=1 waves are expected to vary only by $\sim$1$\%$.  However, \citet{All16} find a maximum period for PSO J318.5-22 of 10.2 hours from $v sin i$ measurements, ruling out the possibility of a double-peaked lightcurve, thus we expect the 8.6$\pm$0.1 hour rotational period derived from our single sinusoid MCMC fits to be accurate in this case.    

\begin{deluxetable}{|l|lccc|}
%\begin{center}
%\begin{table}
%\begin{tabular}{ l | l | c | c | c}
\tablecaption{MCMC fit results for Spitzer variability monitoring. Peak-to-median variability amplitudes are reported here. \label{tab:Spitzermcmc}}
\tablehead{
\colhead{model} & \colhead{parameter} & \colhead{best value} & \colhead{68$\%$} & \colhead{95$\%$}}
\startdata
sine & amplitude & 1.69$\%$ & 0.03$\%$ & 0.07$\%$ \\
 & period & 8.61 hours & 0.06 hours & 0.11 hours \\ 
 & phase & -10.5$^{\circ}$ & 2.4$^{\circ}$ & 4.8$^{\circ}$ \\ \hline
%sine & phase & 
sine+slope & amplitude & 1.81$\%$ & 0.04$\%$ & 0.07$\%$ \\
 & period & 8.63 hours & 0.05 hours & 0.10 hours \\ 
 & phase & -8.4$^{\circ}$ & 2.2$^{\circ}$ & 4.3$^{\circ}$ \\
 & slope & 0.05$\%$ & 0.005$\%$ & 0.01$\%$ \\ \hline
\enddata
%sine+slope & phase &
\end{deluxetable}

\subsection{Inclination}

\begin{figure*}
% For example, with the graphicx style use
\includegraphics[scale=0.24]{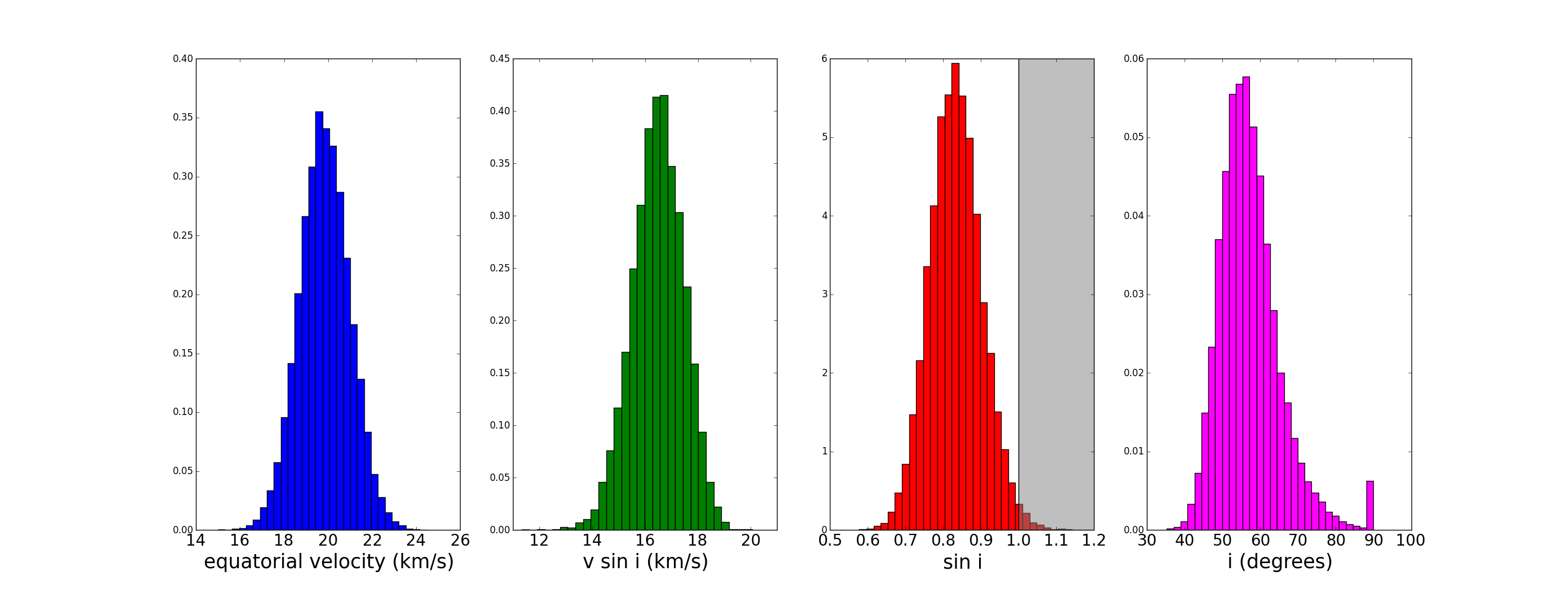}
\caption{Left: Gaussian distribution in equatorial velocity derived from our measured period and radius estimates from \citep{All16}.  Center left: $v sin i$ distribution from fits of the high resolution spectrum of PSO J318.5-22 from \citep{All16}. Center right: $sin i$ distribution.  The shaded gray rectangle indicates values of $sin i$ above 1, which are unphysical and are a result of our adopted uncertainties in radius, period, and $v sin i$ for this object.  Right: inclination distribution.  Unphysical values of $sin i$ have been pinned to 1 here (i.e. 90$^{\circ}$ inclination). 
}
\label{fig:inclination}       % Give a unique label
\end{figure*}

\citet{All16} constrain the inclination of PSO J318.5-22 to $>$29$^{\circ}$, based on their measured $v sin i$, reasonable estimates of the radius of PSO J318.5-22 from on evolutionary models, and a lower limit on the period of $\sim$5 hours from \citet{Bil15}.  With our high-precision measurement of the period from Spitzer observations, we can now directly measure the inclination for this object.  
%Adopting $v sin i$=17.5 $km~s^{-1}$ \citep{All16}, a radius of 1.4 $R_{Jup}$ \citep{All16}, and a period of 8.6 hours from this work implies an equatorial velocity of 19.8 $km~s^{-1}$ and an inclination of 61.8$^{\circ}$.  
Using Monte Carlo methods to account for uncertainties in $v sin i$, radius, and period, we drew 30000 samples from the $v sin i$ distribution found by \citet{All16} and 
Gaussian distributions centered at radius = 1.4$\pm$0.08 $R_{Jup}$ \citep[the~mean~and~standard~deviation~of~the~radius~values~found~in][]{All16}  and period = 8.61$\pm$0.06 hours.  The resulting distributions for equatorial velocity (derived from radius and our measured rotation period), $v sin i$, $sin i$, and inclination are presented in Fig.~\ref{fig:inclination}.  We adopt a value for $sin i$ of 0.83$\pm$0.07, from the median and standard deviation of our $sin i$ distribution.  Propagating errors in the normal way, we find $i = 56.2\pm 7.2^{\circ}$.  Values of $sin i >$ 1 are unphysical and are a result of our uncertainties in measuring radius and $v sin i$.  We have tried to mitigate this issue in two manners: 1) discarding all values of $sin i >$ 1, we find an inclination of 56.1$\pm$7.4$^{\circ}$ (median and standard deviation of remaining values) and 2) pinning all $sin i$ values greater than 1 to 1 (as such a value does imply a high inclination), we find an inclination of 56.2$\pm 8.1^{\circ}$.  All methods provide consistent values for inclination.

\subsection{Amplitude and phase shifts from the HST light curves}

We adopted a similar Markov-Chain Monte Carlo methodology to interpret the HST light curves.  Since the HST observation does not cover a full period, we fixed the period to 8.6 hours, as determined from the Spitzer light-curve.  The MCMC code was run for each of the bands that we extracted lightcurves; posterior pdfs for each band are presented in Appendix~\ref{sec:appendix}.  Again, we adopt the 50$\%$-quantile value as the best value for each parameter.  For each synthesized light curve, a set of 100 samples drawn from the posterior pdf is overplotted on the unbinned HST lightcurve for that band in Fig.~\ref{fig:HST_mcmc_samples}.  We find very little covariance between phase and amplitude in the posterior pdfs, suggesting the fits are robust, at least for the purpose of estimating amplitudes and phase shifts.  Amplitudes (peak to median value) and phases measured for each band from the MCMC fits, as well as phase offsets relative to the Spitzer lightcurve, are presented in Table~\ref{tab:HSTmcmc}.

\begin{deluxetable}{|l|lcc|lcc|lcc|}
\tablecaption{Peak-to-Median Variability Amplitudes, HST phases, and phase offsets relative to the Spitzer lightcurve for synthesized HST band lighcurves \label{tab:HSTmcmc}}
\tablehead{
\colhead{band} & \colhead{amplitude} & \colhead{68$\%$} & \colhead{95$\%$} &  \colhead{HST phase} & \colhead{68$\%$} & \colhead{95$\%$} & \colhead{phase offset} & \colhead{68$\%$} & \colhead{95$\%$}
}
\startdata
white light  & 2.51$\%$ & 0.11$\%$ & 0.23$\%$ & 196.4$^{\circ}$ & 2.4$^{\circ}$ & 4.8$^{\circ}$  & 206.9$^{\circ}$ & 3.1$^{\circ}$ & 6.1$^{\circ}$ \\
(1.07 - 1.67 $\mu$m) & & & & & & & & & \\ \hline
2MASS J & 2.92$\%$ & 0.16$\%$ & 0.32$\%$ & 193.2$^{\circ}$ & 2.9$^{\circ}$ & 5.7$^{\circ}$ & 203.7$^{\circ}$ & 3.4$^{\circ}$ & 6.8$^{\circ}$ \\ \hline
2MASS H & 2.28$\%$ & 0.13$\%$ & 0.25$\%$ & 198.1$^{\circ}$ & 2.9$^{\circ}$ & 5.7$^{\circ}$ & 208.6$^{\circ}$ & 3.4$^{\circ}$ & 6.8$^{\circ}$ \\ \hline
water band  & 2.38$\%$ & 0.19$\%$ & 0.37$\%$ & 199.9$^{\circ}$ & 3.9$^{\circ}$ & 7.7$^{\circ}$ & 210.4$^{\circ}$ & 3.4$^{\circ}$ & 8.7$^{\circ}$\\
(1.34 - 1.44 $\mu$m) & & & & & & & & & \\ \hline
methane band  & 2.20$\%$ & 0.16$\%$ & 0.31$\%$ & 198.5$^{\circ}$ & 3.6$^{\circ}$ & 7.2$^{\circ}$ & 209.0$^{\circ}$ & 4.1$^{\circ}$ & 8.1$^{\circ}$\\ 
(1.60 - 1.67 $\mu$m) & & & & & & & & & \\ \hline
\enddata
\end{deluxetable}

\subsection{Model fits to HST Spectroscopy \label{sec:spec}}

We fit the median, orbit 1 (close to minimum), orbit 3 (bracketing maximum flux), and orbit 5 (close to minimum) spectra of PSO J318.5-22 with three sets of atmospheric models utilizing very different treatments of cloud parameters: (1) the ExoREM models, which focus specifically on low-surface gravity atmospheres in the cloudy, clear, and partly cloudy cases \citep[][Charnay~et~al.~in~prep]{Baud15}, (2) the BT-Settl models, which explore a wide range of dust species grain formation in the presence of hydrodynamical mixing \citep{All11}, and (3) the thick cloud models of \citet{Mad11} (henceforth M11).

\subsubsection{ExoREM model fits}

We compared the grid of ExoREM models to the median spectrum of PSO J318.5-22 through a $\chi^{2}$ minimization.  Unlike most other 1-d models, ExoREM enables the modeling of clear, patchy, and fully cloudy atmospheres, parameterized according to the $f_{c}$ parameter, where $f_{c}$ = 0 is a clear atmosphere and $f_{c}$ = 1 is a fully cloud-covered atmosphere.  As a consequence of the fits, we renormalized each synthetic spectrum by a dilution factor $R^{2}/d^{2}$ which minimizes the $\chi^{2}$, where $R$ is the radius and $d$ the distance to the target.  The fit was performed independently for each value of $f_{c}$ and we compared the best $\chi^{2}$ a posteriori. We considered a fit with $R$ left unconstrained, and another one with the radius varying in the interval $R$ = 1.4$\pm$0.08 $R_{Jup}$ \citep[derived~from~evolutionary~model~fits~assuming~an~age~of~23$\pm$3~Myr,][]{All16}. 
We adopt a distance of 22.2$\pm$0.8 pc (parallax of 45.1$\pm$1.7 mas) from \citet{Liu16}.
%That interval corresponds to DUSTY evolutionary models predictions (Chabrier et al. 2000) for the luminosity of the object (Liu et al. 2013) and an age interval of 10-30 Myr encompassing broadly that of the $\beta$ Pictoris moving group. 

The data are systematically best represented by the ExoREM models with full cloud cover. When $R$ is left unconstrained, we find a best fit for $\mathrm{T_{eff}=1250}$K, log g=3.4 dex, M/H=+0.5 dex, and $R$ =1.12~R$_{Jup}$. When an a-priori range on the radius is given, the spectrum of PSO J318.5-22 is best reproduced for  T$_\mathrm{eff}$=1150 K, log g=3.3 dex, M/H=0.0 dex, and $R$=1.39~R$_\mathrm{Jup}$.  We show the solution in Fig.~\ref{fig:EXORemfits} along with a $\chi^{2}$ map for the solar-metallicity models. The model reproduces the spectral slope and the object NIR brightness simultaneously, but it fails to reproduce the detailed morphology of the H band and the strength of the water-absorption at 1.3-1.5 $\mu$m. The variation of the spectral features of PSO J318.5-22 are within the error bars of the median spectrum, thus the fitting solutions are identical when our various HST spectra of PSO J318.5-22 are considered.

\subsubsection{BT-Settl and M11 model fits \label{sec:BT-Settl_M11}}

For the BT-Settl and M11 models, fits were performed with a Levenberg-Marquardt least square fitter $(numpy.optimize.curvefit)$, after binning the model spectra to the same resolution and sampling as the HST spectra.  Best fit models are shown in Fig.~\ref{fig:BTSettlM11fits}.  While both of these models produce spectra which are roughly qualitatively similar to our observed spectra, there are notable differences, namely, the models do not reproduce the steepness of the observed spectral slope from 1.2 to 1.35 $\mu$m or from 1.4 to 1.7 $\mu$m.  As the difference between the model spectra and our observed spectra were larger than the difference between observed spectra from different orbits, we found that a single model spectrum fit best all the single orbit spectra as well as the median spectrum.  In other words, we were unable to describe the changes seen in the observed spectra as a function of time by fitting different model spectra with varying $T_{eff}$ and surface gravities.  The 1D static models used here, however, are essentially averages over the entire visible surface area of the object and over multiple rotational periods, so it is is not surprising that we find similar fits for both single orbit and median spectra.

The best fit BT-Settl model had $T_\mathrm{eff}$=1600 K and $log(g)=3.5$, although a range of models with $T_\mathrm{eff}=1500-1700 K$ and $log(g)=3-5$ fit the spectra nearly as well.  The $T_\mathrm{eff}$=1600 K and $log(g)=3.5$ best fit is driven by the fitting algorithm's attempt to fit the spectral slope in H band.  In contrast, a $T_\mathrm{eff}$=1700 K, $log(g)=5.0$ fits the J band slope better, at the expense of a very poor H band fit \citep[similar~to~what~was~found~for~HR~8799de~by][]{Bon16}.  
%This behavior is consistent with similar model fits for HR 8799de presented in \citep[][]{Bon16}.  
These results are consistent with the best fits reported in \citet{Liu13} for the IRTF SpeX spectrum ($T_\mathrm{eff}=1400-1600 K$ and $log(g)=4.0 - 4.5~dex$).  As noted in 
\citet{Liu13}, this set of atmospheric model spectra fits yield $T_\mathrm{eff}$ values that are significantly higher than those obtained from evolutionary model fits to photometry.  For the \citet{Mad11} models, the best fits were obtained for model A (thick clouds) with 60 $\mu$m grains, $T_\mathrm{eff}=1100-1200 K$ and $log(g)=3.75 - 4.25~dex$, consistent with the IRTF SpeX spectrum fits (model A, 60 $\mu$m grains,$T_\mathrm{eff}=1100 K$ and $log(g)=4.0~dex$) reported in \citet{Bil15}.

As with the ExoREM model fits, we renormalized each synthetic spectrum by a dilution factor $R^{2}/d^{2}$ which minimizes the $\chi^{2}$, where $R$ is the radius and d the distance to the target. We again adopt a distance of 22.2$\pm$0.8 pc (parallax of 45.1$\pm$1.7 mas) from \citet{Liu16}. Thus, from the dilution factor obtained from the model fit, we estimate the radius $R$ of PSO J318.5-22.  This results in an unphysically small radius estimate of $\sim$0.7-0.8 R$_{Jup}$ using the BT-Settl models for this young, low surface gravity object.  \citet{Liu13} obtained a similar result fitting this same model set to a low-resolution near-IR spectrum; in contrast, adopting an age of 23$\pm$3 Myr, \citet{All16} find $T_\mathrm{eff}$ = 1100--1200 K and radius values of 1.34 -- 1.46 $R_{Jup}$ using a variety of evolutionary model grids with and without clouds.  The observed luminosity of a (very roughly blackbody) object is governed by the temperature and the radius; the high temperature fit by the BT-Settl models has thus necessitated an unphysically small fit to the radius to produce the observed luminosity.  For the M11 models, with a cooler fitted temperature of 1100-1200 K, we find radius estimates of $\sim$1.0-1.3 R$_{Jup}$, roughly consistent with our ExoREM model fits with radius left as a free parameter, but still somewhat smaller than estimates based on evolutionary models \citep{Liu13,All16}.

\section{Discussion}

\subsection{Amplitudes \label{sec:amplitudes}}

\begin{figure*}
\includegraphics[scale=0.4]{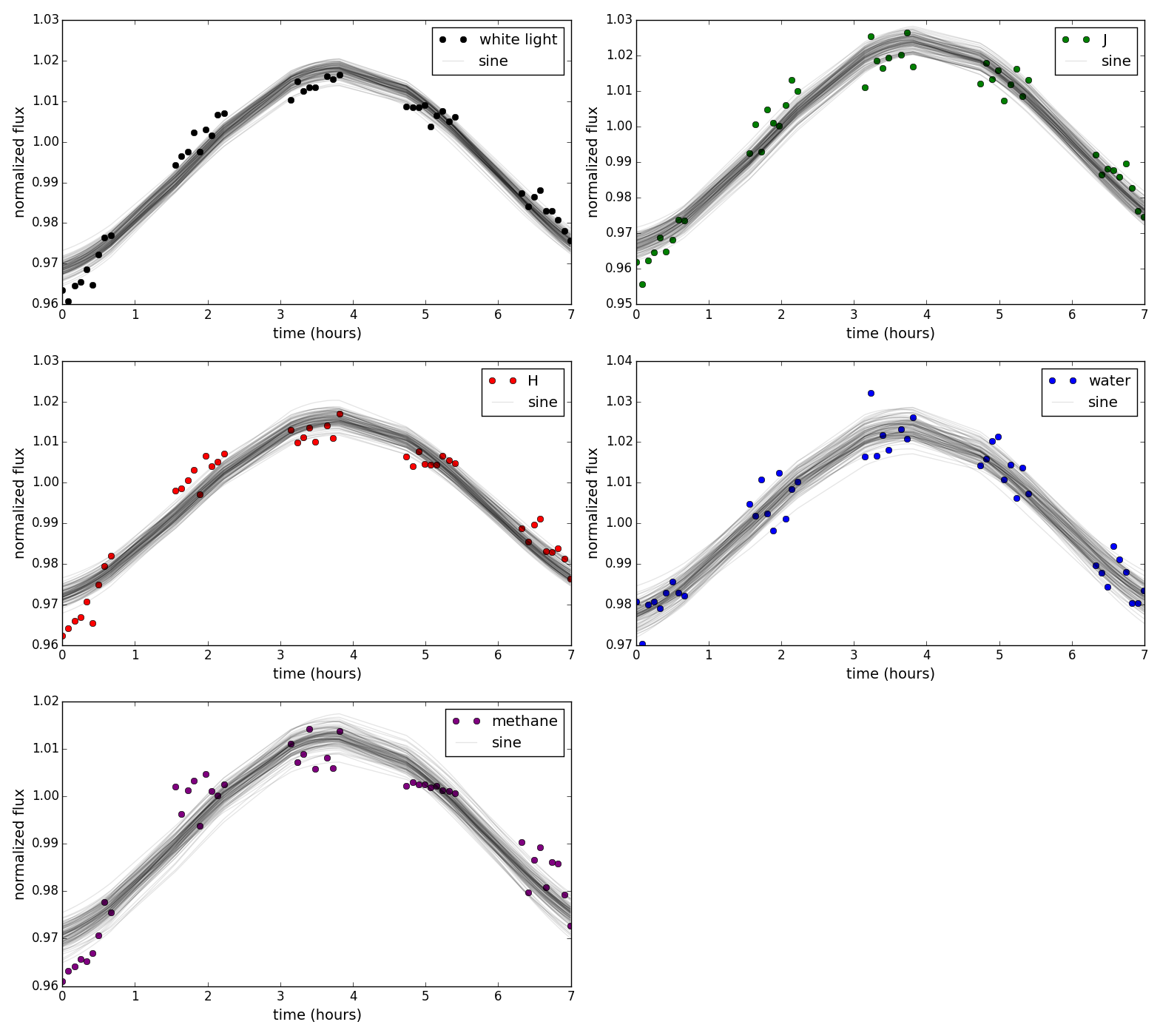}
\caption{100 sample model lightcurves drawn from our sinusoid (black) MCMC fits to each synthesized HST lightcurve (white light, 2MASS J, 2MASS H, water, and methane).  All lightcurves except the water lightcurve show a significant deviation from the sinusoidal fits in the first 30 minutes of the observation; given that the water lightcurve appears sinusoidal, this is not a result of the ramp effect, which should affect all wavelengths equally.
}
\label{fig:HST_mcmc_samples}       % Give a unique label
\end{figure*}

\begin{figure*}
% For example, with the graphicx style use
\includegraphics[scale=0.8]{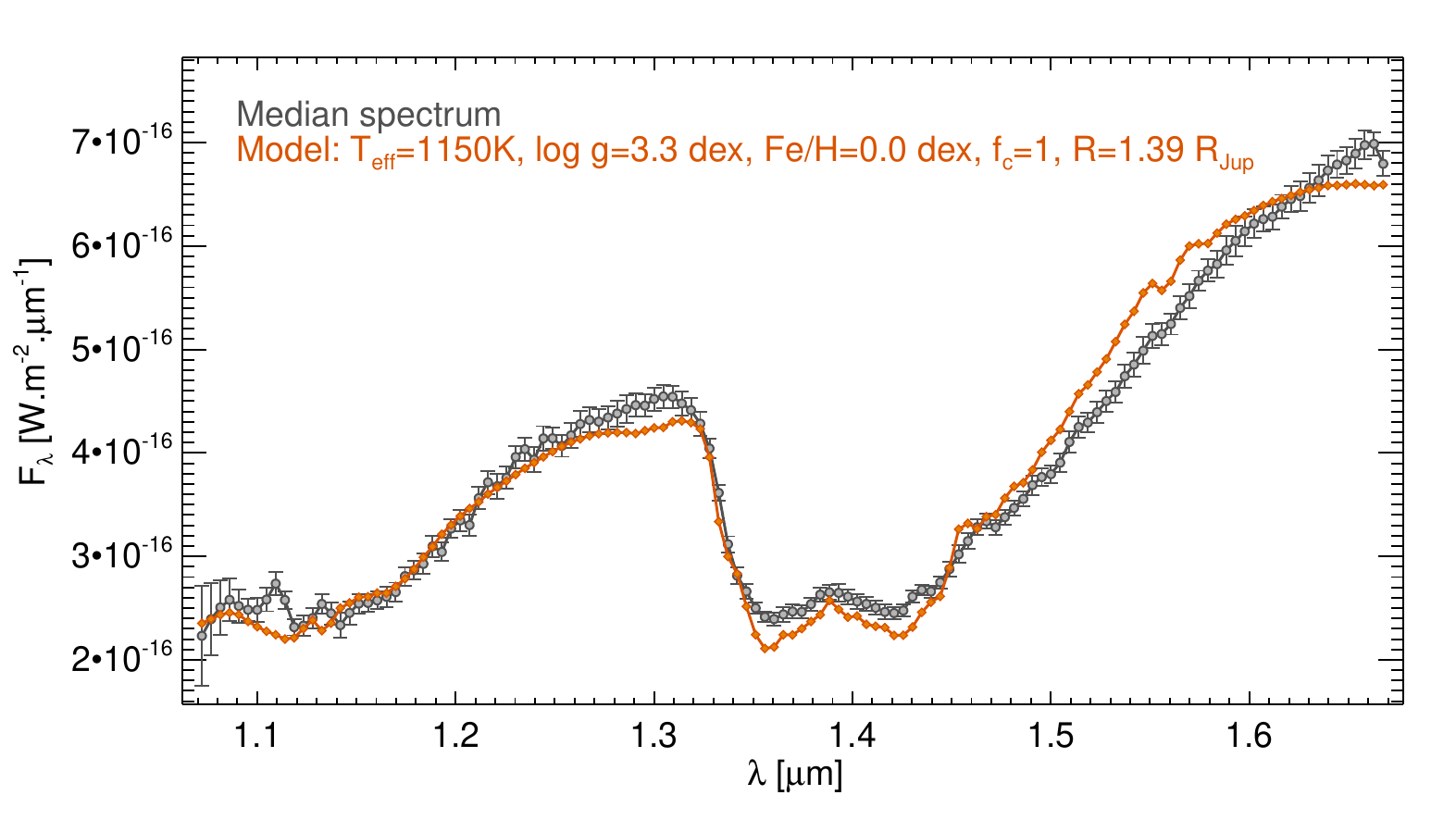}
\includegraphics[scale=0.8]{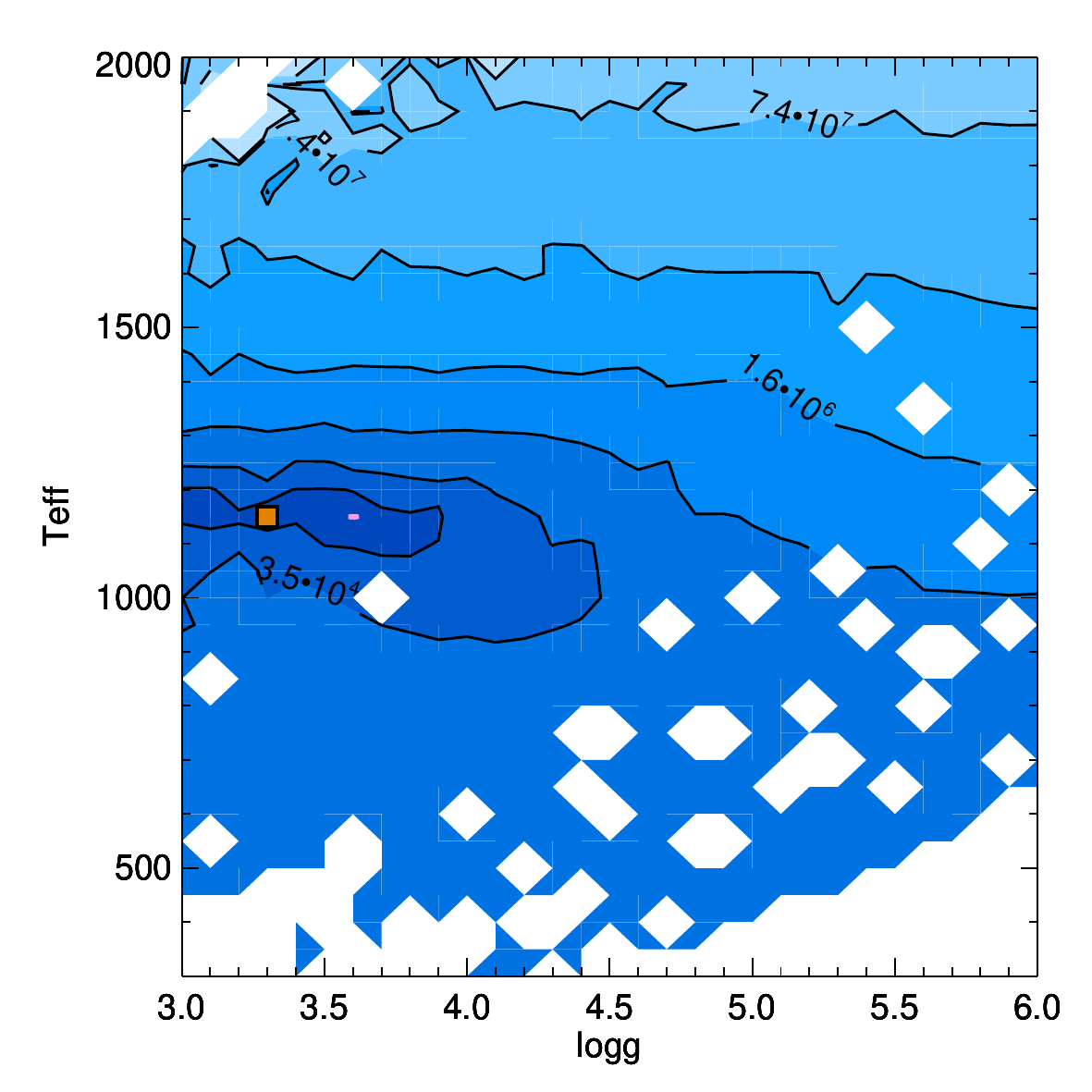}
\caption{Top: Comparison of the HST median spectrum of PSO J318.5-22 to the ExoREM models when the radius is allowed to vary in the range  1.29 - 1.35 $R_{Jup}$. Bottom: $\chi^{2}$ map for the solar-metallicity models with a full cloud cover. The orange square corresponds to the $\chi^{2}$ minimum. 
}
\label{fig:EXORemfits}
\end{figure*}

\begin{figure*}
% For example, with the graphicx style use
\includegraphics[scale=0.7]{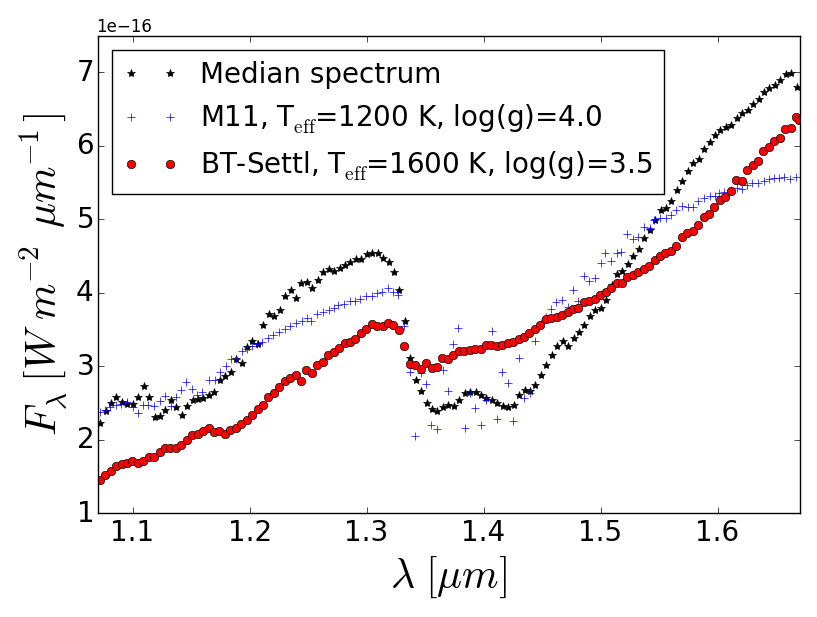}
\caption{Best BT-Settl (red circles) and M11 model (blue crosses) fits overplotted with the HST median spectrum of PSO J318.5-22 (black stars).  These models do not reproduce the steepness of the observed spectral slope from 1.2 to 1.35 $\mu$m or from 1.4 to 1.7 $\mu$m.  
The best fit BT-Settl model had $T_\mathrm{eff}$=1600 K and $log(g)=3.5$, although a range of models with $T_\mathrm{eff}=1500-1700 K$ and $log(g)=3-5$ fit the spectra nearly as well.  The $T_\mathrm{eff}$=1600 K and $log(g)=3.5$ best fit is driven by the fitting algorithm's attempt to fit the spectral slope in H band.  For the \citet{Mad11} models, the best fits were obtained for the model A (thick clouds), 60 $\mu$m grains, $T_\mathrm{eff}=1100-1200 K$ and $log(g)=3.75 - 4.25~dex$.    
}
\label{fig:BTSettlM11fits}
\end{figure*}

Over our 7 hours of HST monitoring, we only captured one clear extremum, a maximum in brightness that occurs in orbit 3.  The minimum value of brightness measured during our time series occurred in orbit 1. However, orbit 1 was the most affected by the ramp effect.  Orbit 5 is also near a minimum of the lightcurve and should not be affected as strongly by the ramp effect.  The ratios of maximum and minimum spectra (taking both orbit 1 and orbit 3 as potential minima) are plotted in Fig.~\ref{fig:HST_min_max_ratio}. The orbit 3 spectrum divided by the orbit 5 spectrum shows a monotonic and relatively small decrease in this ratio as a function of wavelength.  In contrast, for the orbit 3 spectrum divided by the orbit 1 spectrum, the ratio of maximum to minimum spectral flux in the 1.4 $\mu$m water absorption feature is slightly smaller relative to that at adjacent shorter and longer wavelengths, superimposed on a monotonic, small decrease in variability amplitude as a function of increasing wavelength.  The suppression of variability in the water band relative to the adjacent continuum in orbit 1 appears to be robust -- from Fig.~\ref{fig:HST_mcmc_samples}, all lightcurves except the water band lightcurve display a significant deviation from the sinusoidal MCMC fits during Orbit 1.  This is likely not a result of the ramp effect, which should affect all wavelengths equally.  
For the L/T transition objects SIMP 0136 and 2M 2139, \citet{Apa13} and \citet{Yan16} found significantly smaller amplitudes in the water absorption feature relative to both the shorter and longer wavelength continuum, while for 2 mid-L dwarfs, \citet{Yan15} found small but monotonic decreases of amplitude with increasing wavelength.  Our results appear to be a hybrid of these cases, with different behavior observed in different orbits.        

As our HST observations did not cover a full period, we did not measure the full amplitude of variability.  However, we did cover the majority of the period and have a robust period determination from the simultaneous Spitzer observations, so we can use the sinusoidal fits to the lightcurves to estimate amplitudes and phase shifts (relative to the Spitzer lightcurve) for the 5 broadband regions we have considered previously: the full white light spectrum from 1.1 to 1.67 $\mu$m, 2MASS J, 2MASS H, water, and methane.  As noted previously, this method appears to actually slightly underestimate variability amplitude, as most of the synthesized lightcurves show some deviation from the sinusoidal fits during orbit 1.  Measured amplitude and phase shift for each of the broadband regions we considered are plotted as a function of wavelength in Fig.~\ref{fig:HST_lambda_phase_amplitude}.  Similar to the divided spectra for orbit 3 and orbit 5, amplitude appears to generally decrease as a function of increasing wavelength, with the sharpest break between $J$ and 1.4 $\mu$m water band.   %The amplitude does not appear to be significantly smaller in the water band compared, as had been found for the L/T transition dwarfs SIMP 0136 and 2M 2139 \citet{Apa13}.  Instead, the dependence on amplitude appears to be similar to that found for mid-L dwarfs by \citet{Yan15} -- i.e. generally decreasing at longer wavelengths across the near-IR.  

The mid-IR Spitzer [4.5 $\mu$m] lightcurve follows the same trend of decreasing amplitude with longer wavelength, with a peak-to-trough amplitude of $\sim$3.4$\%$ vs. 4.4 - 5.8$\%$ for the near-IR bands.  For field brown dwarfs, near-IR variability is generally found to have a significantly higher amplitude than mid-IR variability if both are present.  It is notable that the mid-IR variability amplitude for PSO J318.5-22 is so similar to its near-IR variability amplitude and is in fact one of the highest variability amplitudes ever measured in the mid-IR for a brown dwarf or planetary mass object!  The highest amplitude variable from \citet{Met15} is the T6 dwarf 2MASS~J22282889-4310262, with a 3.6 $\mu$m variability amplitude of 4.6$\pm$0.2$\%$, but most of the variables in their sample have amplitudes of $<$2$\%$.  
This high amplitude variability in the mid-IR may be the effect of low surface gravity on the vertical structure of such an atmosphere.  Low surface gravity allows cloud species to potentially extend up to lower pressures and higher altitudes compared to the high surface gravity case \citep{Mar12}.  In general for brown dwarfs and free-floating planetary mass objects, the photosphere in the mid-IR is at lower pressures and higher amplitudes than the photosphere in the near-IR photosphere \citep[see~e.g.][]{Mar12,Bil13}.  Thus, the extension of clouds up to lower pressure regions increases the chance of heterogeneous cloud opacity (and hence variability) at the low pressure levels probed by mid-IR observations.

The viewing inclination of a given object will significantly affect its variability properties \citep[see~e.g.][]{Vos17}.  Presuming that surface features that generate variability are primarily equatorial, the same object observed at high inclination will appear to have a higher variability amplitude than if viewed at lower inclination.  Additionally, \citet[][]{Vos17} find that J-band variability is more affected by inclination angle than mid-IR variability.  As J-band observations generally probe a deeper part of the atmosphere than mid-IR observations \citep[][]{Bil13b,Yan16,Vos17}, \citet[][]{Vos17} propose that this effect may be due to the increased atmospheric path length of J-band flux at lower inclinations.  We measure a relatively high inclination for PSO J318.5-22 of 56$\pm$8$^{\circ}$, thus, we are observing close to the full amplitude in each band.    
%suggesting that inclination is not preferentially affecting observed amplitude in one waveband vs. the other. 

\subsection{Phase Shifts \label{sec:phaseshifts}}

We measure a phase shift between the Spitzer lightcurve and HST lightcurves of $\sim$200$^{\circ}$.  At a lower significance, we find a $\sim$6-7$^{\circ}$ phase shift between the J-band lightcurve and the other near-IR narrow band lightcurves.  Similar results have been found for older, high-surface gravity brown dwarfs -- phase shifts between the near-IR and mid-IR lightcurves may be quite common for these objects.  The first such phase shift was found by \citet{Bue12} for the T6 2MASS~J22282889-4310262 and \citet{Yan16} recently found significant phase shifts between the mid- and near-IR for 4 high-surface gravity brown dwarfs (including 2MASS~2228).  In 3 out of 4 cases from \citet{Yan16}, the measured phase shift between near-IR and mid-IR is nearly $\sim$180$^{\circ}$, ranging from 150-210$^{\circ}$ (see their Figs. 19-22), very similar to the $\sim$200$^{\circ}$ phase shift reported here for PSO J318.5-22.  
Phase shifts within the near-IR spectral bands are less common \citep{Bue12,Bil13,Yan16} but have been reported now at multiple epochs for the T6 2MASS~J22282889-4310262 \citep{Bue12,Yan16}.  For PSO J318.5-22, all the near-IR bands we consider agree in phase at the 2$-\sigma$ level; at the 1$-\sigma$ level, 2MASS J is shifted by $\sim$6 degrees relative to the other near-IR bands.  However, given that we have monitored less than one rotation period with HST and have also assumed a sinusoidal light curve shape, this phase shift is still within the errors expected from our sinusoidal model fitting.  Observed phase shifts have generally been interpreted as different "top-of-atmosphere" locations at different wavelengths -- near-IR generally probes deeper in the atmosphere than mid-IR \citep{Bue12,Bil13,Yan16}.  In other words, the source of inhomogeneity which drives near-IR variability is located at a higher pressure level deeper in the atmosphere than the source of inhomogeneity driving mid-IR variability at a lower pressure level.

\subsection{Principal Component Analysis of HST spectra}

Following the method of \citet{Apa13}, we perform principal component analysis (henceforth PCA) to determine how many spectral components drive the variability for PSO J318.5-22.  
We expect the observed variability to be driven by the rotation in and out of view of regions with differing spectra. PCA identifies the smallest set of independent spectra that account for the majority of the observed variability.  Taking each of the 49 spectra in our spectral sequence as a dimension and subtracting off the mean value for each spectrum (as required for PCA), we calculate the 49$\times$49 covariance matrix between each spectrum using the $numpy.cov$ function in python.  We then used $numpy.linalg.eig$ to determine eigenspectra and eigenvalues.  Sorting on eigenvalues, the principal spectral component accounted for 99.6\% of the observed variability, with the second component contributing 0.1$\%$ and the third component contributing 0.07$\%$.  This is similar to results for other variable objects -- \citet{Apa13} find that the principal spectral component accounted for 99.6$\%$ and 99.7$\%$ of the variability for 2M2139 and SIMP 0136 respectively.  They argue that this implies that only two types of surface patches are required to explain the observed variability for these objects. However, subtracting off the mean essentially means that we remove any gray variation -- any shifting of the entire spectrum by a constant value.  From Fig.~\ref{fig:HST_min_max_ratio} and Fig.~\ref{fig:HST_lambda_phase_amplitude}, the observed variability does not possess a strong color component, as variability amplitude changes monotonically and slowly with wavelength over the 1.07-1.67 $\mu$m spectral range of the WFC3 G141 grism -- for instance, the variability amplitude in the methane band is roughly 75$\%$ of that in the $J$ band.  Thus, removing the mean for the PCA actually removes most of the observed variability.  Hence it is not surprising that the principal spectral component (which closely resembles the median spectrum) encompasses the vast majority of the power in the time-series.   

\subsection{Possible Sources of Variability}

The key variables for constraining variability properties in both field brown dwarfs and
young exoplanets are spectral type (which correlates with T$_\mathrm{eff}$) and surface
gravity. High temperature objects (spectral types from early to mid-L) may have variability
from cloud features as well as magnetic activity (e.g. star-spots and aurorae); lower temperature objects ($>$L5) are assumed to have atmospheres too cool to produce magnetic activity and are presumed to have entirely cloud-based variability \citep{Gel02}. The variability of older, high-surface gravity
field dwarfs across the full L-T spectral has been studied in
detail spectroscopically \citep{Bue15a,Bue15b,Apa13,Bue12,Yan15}.
However, only three young, low-surface gravity objects have HST spectral variability monitoring to date -- SIMP0136 \citep{Apa13}, W0047 \citep{Lew16},
and the observations presented herein for PSO J318.5-22.  Here we consider a
number of drivers of variability suggested in the literature for both high and low surface gravity objects (specifically patchy cloud features, hot spots, high-level hazes, thermochemical instabilities, and magnetically-driven aurorae), to determine if they can describe the variability properties observed for PSO J318.5-22.

%Surface gravity is expected
%to strongly affect variability properties, which we attempt to quantify for the case of 
%PSO J318.5-22 relative to the well-studied cohort of high-surface gravity brown dwarfs.  

Considering patchy salt and sulfide clouds as well as hot spot models (heating at a specific pressure level) for high-surface gravity objects with T$_\mathrm{eff}>$375 K, \citet{Mor14b} find that variability due to patchy clouds should drive high amplitude variability across a wide spectral range while variability due to hot spots should produce larger variability amplitudes within absorption features relative to continuum wavelengths.   As noted in Section~\ref{sec:phaseshifts}, for PSO J318.5-22, we find a smooth decrease of variability amplitude as a function of increasing wavelength across the 1.07-1.67 $\mu$m spectral range.
%, with a potential slight decrease in variability amplitude in the 1.4 $\mu$m water absorption feature in orbit 1.  
For the 1.4 $\mu$m water absorption feature and the 1.6$\mu$m methane absorption feature covered by the HST WFC3 grism, we find similar or slightly smaller variability amplitudes relative to the adjacent continuum.
%, thus suggesting cloud features as the driver of the observed variability.
The lack of stronger variability within absorption features compared to continuum wavelengths
for PSO J318.5-22 implies that inhomogeneous cloud features (thick and thin clouds or a haze layer over a thick cloud surface) is likely to be a major driver of the observed variability.
 
\begin{figure*}
% For example, with the graphicx style use
\includegraphics[scale=0.7]{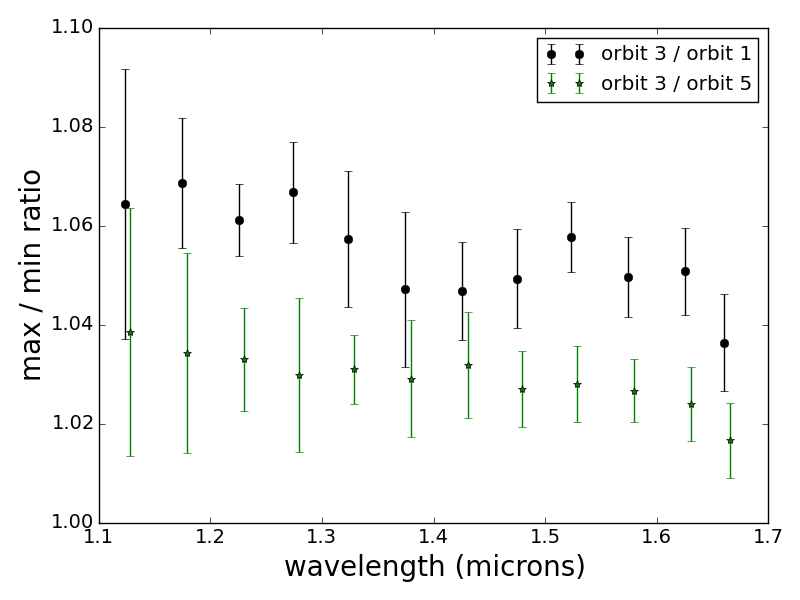}
\caption{Ratio of maximum and minimum PSO J318.5-22 HST spectra, binned by 0.05 $\mu$m.  The minimum value of brightness measured during our time series occurred in orbit 1. However, orbit 1 was the most affected by the ramp effect.  Orbit 5 is also near a minimum of the lightcurve and should not be affected as strongly by the ramp effect.  Thus, we plot here two max / min spectral ratios: orbit 3 divided by orbit 1 and orbit 3 divided by orbit 5.  The spectral ratio for orbit 3 divided by orbit 5 has been offset slightly in wavelength for clarity.  As our HST observations did not cover a full period, these are lower limits on the full amplitude.  
}
\label{fig:HST_min_max_ratio}       % Give a unique label
\end{figure*}

The 1.4 $\mu$m water absorption feature can potentially provide a useful diagnostic of cloud-driven variability mechanisms.  This feature is available from space with HST but is very difficult to observe from the ground.  In the case of variability from high-level hazes \citep[as~observed~for~two~mid-L~field~dwarfs~by][]{Yan15}, 
we expect similar variability amplitudes both with and within the water absorption feature, 
if the high-level hazes driving the variability are located at a very low pressure, high altitude in the atmosphere where the water opacity is negligible.
In the case of variability due to inhomogeneous thin and thick clouds, we expect the variability amplitude to be notably different in the water absorption feature relative to adjacent non-absorbed wavelengths.
For instance, \citet{Apa13} found variability to be suppressed at 1.4 $\mu$m relative to the adjacent continuum for two highly variable L/T transition brown dwarfs. 
However, as noted in Section~\ref{sec:phaseshifts} and Fig.~\ref{fig:HST_min_max_ratio}, 
PSO J318.5-22 displays both behaviors in different orbits!  Thus, in this case, the amplitude inside and outside the 1.4 $\mu$m water absorption feature does not clearly distinguish between the cases of high-level hazes vs. inhomogeneous thin and thick clouds.
%is similar to the two mid-Ls observed by \citet{Yan15}
%in that it does not show a notable change in variability amplitude in the water absorption feature relative to nearby continuum wavelengths; however, 
%The variability 
%amplitudes found for PSO J318.5-22 are significantly higher overall than those for the two mid-Ls from \citet{Yan16} and are 
%in fact similar to the variability amplitudes found for the most highly variable L/T transition field dwarfs.  This suggests significant top-of-atmosphere inhomogeneities in very low-pressure, dry parts of this atmosphere.

\begin{figure*}
% For example, with the graphicx style use
\includegraphics[scale=0.68]{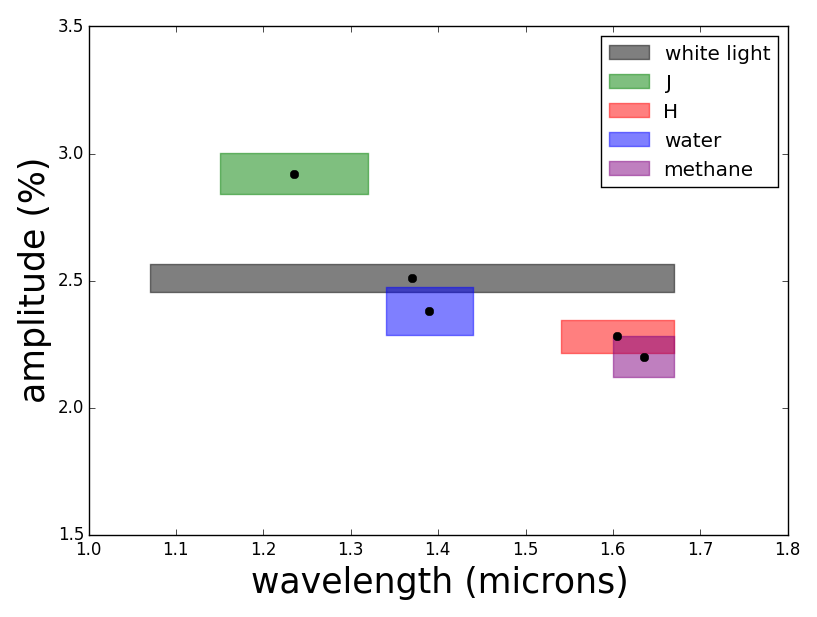}
\includegraphics[scale=0.68
]{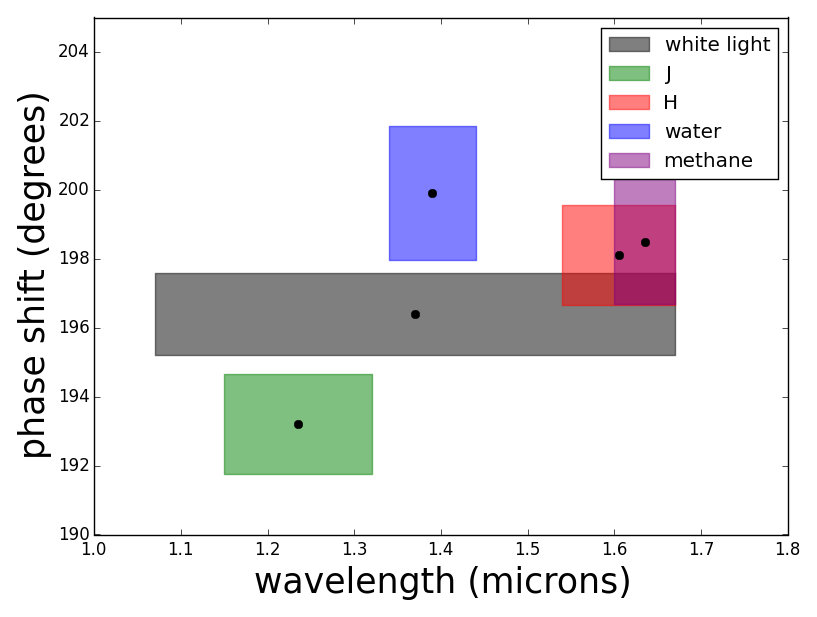}
\caption{Top: Wavelength vs. measured variability amplitude for HST synthesized lightcurves.  Shaded boxes gives the passband used on the wavelength axis and the 1-$\sigma$ error on the amplitude axis.  Amplitude appears to decrease from shorter to longer wavelengths. 
Bottom: Wavelength vs. measured phase relative to the Spitzer channel 2 lightcurve.  Shaded boxes gives the passband used on the wavelength axis and the 1-$\sigma$ error on the phase shift axis.  Phase shifts across each of the synthesized bandpasses agree at the 2-$\sigma$ level; J band is phase shifted by $\sim$6$^{\circ}$ relative to the other near-IR bands.   
}
\label{fig:HST_lambda_phase_amplitude}       % Give a unique label
\end{figure*}

However, recent work has suggested that clouds may not be necessary to model the spectra \citet{Tre16,Trem17} -- or variability -- of L and T spectral type objects.  \citet{Tre16} have recently produced cloud-free models of L and T type brown dwarf atmospheres, successfully modeling the red colors of L dwarfs as well as the T dwarf J band brightening and re-emergence of the FeH absorption feature using additional convection from to thermo-chemical instabilities (in the CO / CH$_4$ transition in the case of the L/T boundary).  They suggest that turbulence produced by CO or temperature fluctuations across the CO / CH$_4$ may be a driver of the observed variability of brown dwarfs, in particular, that the inhomogeneous top-of-atmosphere structure mapped via Doppler imaging for the L/T transition brown dwarf Luhman 16B \citep{Cro14a} may be explained by inhomogeneities in CO vs. CH$_4$ abundance or temperature.  We consider whether this mechanism might drive variability for PSO J318.5-22.
Variability due to abundance variations should drive increased variability amplitudes in the absorption features produced by the species in question.  For this reason, we produced synthesized lightcurves in the 1.6 $\mu$m methane absorption feature -- at least the portion of it which lies within the 1.07-1.67 $\mu$m spectral range of the HST WFC3 G141 grism.  We do not find variability amplitude in the methane band to be significantly enhanced or suppressed relative to the wider 2MASS H band (see Fig.~\ref{fig:HST_lambda_phase_amplitude}).  We tentatively suggest the observed variability is not driven by varying CH$_4$ abundance here -- however the full theoretical calculation of expected variability amplitude in methane absorption features due to thermochemical instabilities is not yet available, thus, we await more quantitative theoretical predictions here.
%, but cannot fully rule out variability due to hot-spots caused by thermo-chemical instabilities.  
%However, for the most part, the observed variability of PSO J318.5-22 is more easily described by variations in longitudinal cloud structure, .  
%While \citet{Tre16} argue that clouds are not necessary to model the L/T transition, they also caution that this does not indicate clouds are absent.

While lower temperature objects ($>$L5) are commonly assumed to lack magnetically-driven variability \citep{Gel02}, \citet{Hal15} suggest that this may not always be the case.  \citet{Hal15} find significant phase shifts between radio and various optical bands for the much-hotter nearby M8.5 object LSR J1835+3259, which they interpret as auroral heating.  In particular, electron beams from global auroral current systems feeds energy from the magnetosphere into the atmosphere of this object.  \citet{Hal15} suggest that this mechanism may extend down even to very cool brown dwarfs, driving some of the more extreme examples of weather phenomena in brown dwarfs. They note as well that radio emission has been detected from objects with spectral types as late as T6.5.
From our current dataset, we cannot confirm or refute if this is the case for PSO J318.5-22 -- radio and H-$\alpha$ monitoring would be necessary to do so.  However, more generally, \citet{Mile17} searched for H-$\alpha$ emission for a sample of eight L3-T2 field brown dwarfs, six of which have detections of photometric variability.  The only H-$\alpha$ detection in this sample was from a non-variable T2 dwarf, suggesting that aurorae and other chromospheric activity do not commonly drive variability for L and T spectral type objects.

\subsection{Theoretical consideration of observed amplitudes and phase shifts in the framework of cloud-driven variability}

\begin{figure*}
% For example, with the graphicx style use
\includegraphics[scale=0.4]{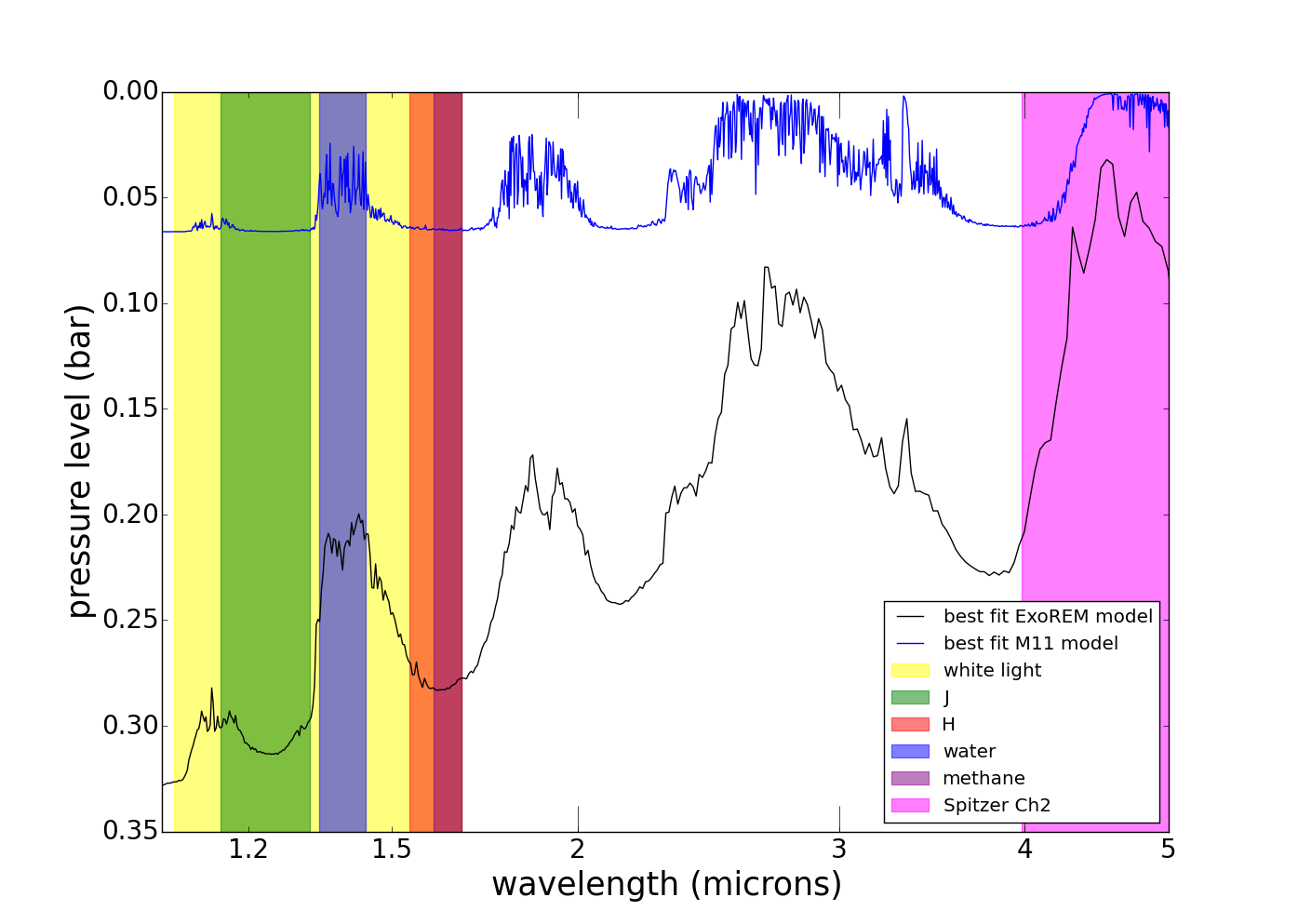}
\caption{Pressure spectra for best-fit ExoREM (black, T$_\mathrm{eff}$=1150 K, log(g)=3.3 dex, M/H=0.0 dex, and R=1.39 R$_{Jup}$) and M11 models (blue, A60, T$_\mathrm{eff}$=1100 K, log(g)=4, solar metallicity).  The bandpasses for the HST synthesized lightcurves and the Spitzer channel 2 lightcurve are shown as shaded boxes.  For both models, mid-IR flux is generated higher in the atmosphere than near-IR flux.}
\label{fig:A60.1100.pressure_levels}       % Give a unique label
\end{figure*}

Assuming cloud-driven variability, what cloud species and cloud geometries are necessary to reproduce our observed amplitudes and phase shifts across the near and mid-IR?
To try to quantify the expected pressure level at which the photosphere is found at a given wavelength for PSO J318.5-22, we considered the best-fit model to our HST median spectrum to identify at what pressure level flux is being emitted at each wavelength.  As our BT-Settl model fit in Section~\ref{sec:BT-Settl_M11} yielded a higher temperature and smaller radius than is consistent with evolutionary model fits to the same object \citep{Liu13,All16}, we consider only our model fits using the M11 models and the ExoREM models for this analysis.  We adopt the ExoREM fully cloudy model with $\mathrm{T_{eff}=1150}$K, log g=3.3 dex, M/H=0.0 dex, and R=1.39$\mathrm{R_{Jup}}$.  In \citet{Bil15}, we found that the SpeX spectrum for PSO J318.5-22 is best fit with the A60, 1100 K, solar metallicity, $log(g)$=4 model from \citet{Mad11} and found similar fits for our HST time-resolved spectra (See Section~\ref{sec:spec}).  As the SpeX spectrum fit is similar to what we find here and covers a wider wavelength range, we adopt this model fit for the M11 models.  Photospheric pressures are provided with the publicly available M11 models; for the ExoREM models, a pressure spectrum was generated by combining the model spectrum with the pressure / temperature profile of the model.  Flux at each wavelength was converted to the equivalent brightness temperature.  We then interpolated using the corresponding pressure / temperature profile to obtain the photospheric pressure level. This method is correct if the source function varies linearly with optical depth.  The resulting "pressure spectra" for both of these models are plotted in Fig.~\ref{fig:A60.1100.pressure_levels}, with the bandwidth for each of our lightcurves overplotted.  While the photospheric pressure level vs. wavelength varies between models, in both cases the mid-IR flux is generated higher in the atmosphere than the near-IR flux.  

\begin{figure*}
% For example, with the graphicx style use
\includegraphics[width=\linewidth]{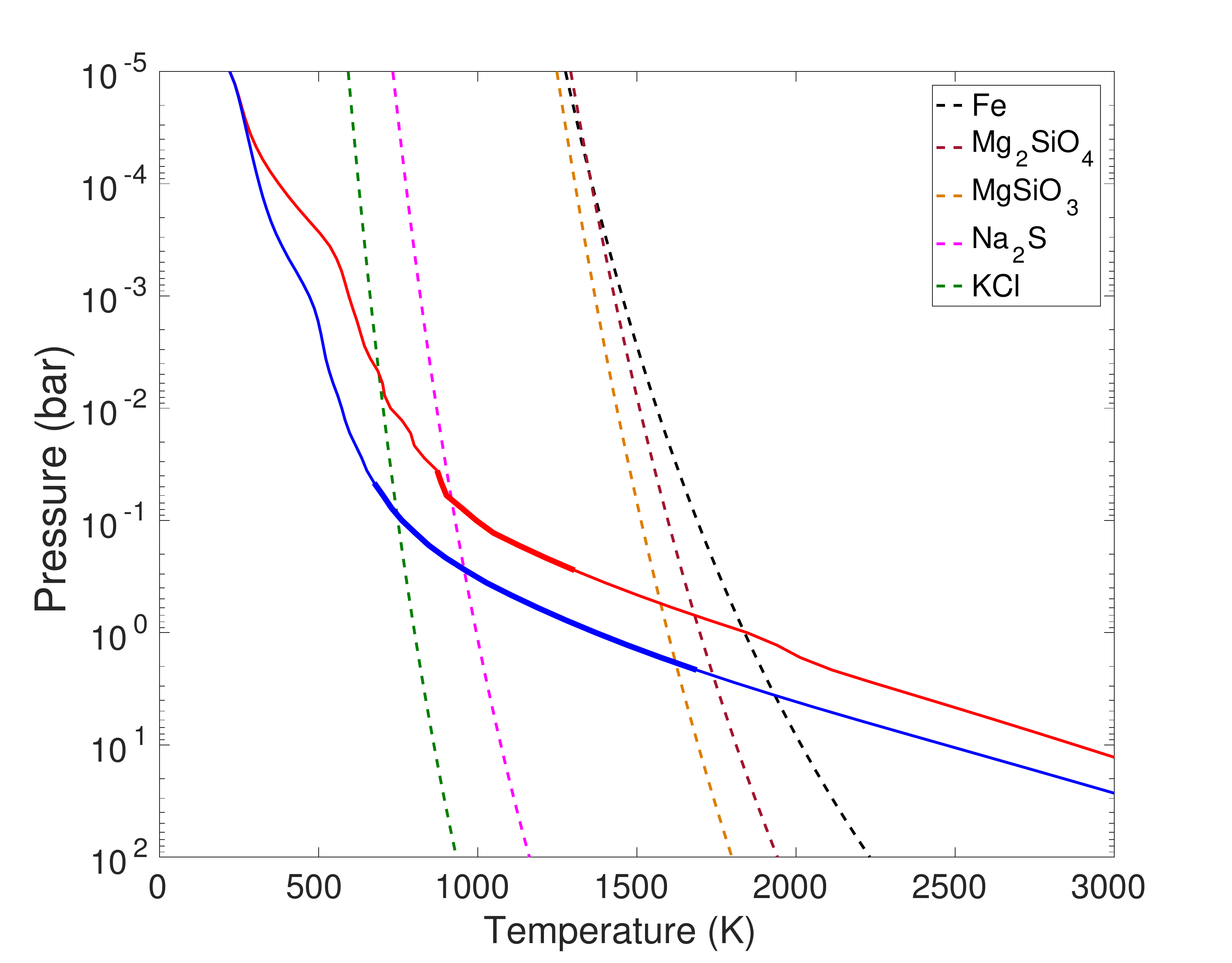}
\caption{Pressure / temperature profile for both our best fit ExoREM cloudy model (red curve) and an equivalent clear model (blue curve) with other parameters unchanged.  Thick lines correspond to the photosphere (computed from 0.6 to 5 $\mu$m) and dashed lines are condensation temperatures for the different clouds present in the model.  The presence of clouds increases the temperature by around 200 K in the photosphere region.}
\label{fig:TP}       % Give a unique label
\end{figure*}

What cloud species are expected to dominate at the respective higher and lower pressures probed by near-IR vs. mid-IR observations?  In Fig.~\ref{fig:TP}, we plot the pressure / temperature profile for both our best fit ExoREM cloudy model and an equivalent clear model with other parameters unchanged.  Thick lines correspond to the photosphere (computed from 0.6 to 5 $\mu$m) and dashed lines are condensation temperatures for the different clouds present in the model.  The presence of clouds (red curve) increases the temperature by around 200 K.  In the cloudy case, the type of cloud which condenses varies according to the pressure / temperature profile and the condensation temperatures for different cloud species.  Silicate and iron clouds form at around 1 bar and are optically thick up to around 0.3 bar at 1~$\mu$m. Thus, for these model atmospheres, silicate and iron clouds form below the photosphere pressures of the 1.4~$\mu$m water band and the 4.5~$\mu$m CO band. In contrast, Na$_2$S clouds form in the upper atmosphere at around 0.06 bar, so above the photosphere pressures of all molecular bands except the 4.5~$\mu$m CO band.
 
\begin{figure*}
% For example, with the graphicx style use
%\includegraphics[scale=0.4]{p_photosphere_1150K_g33b.eps}
\includegraphics[width=\linewidth]{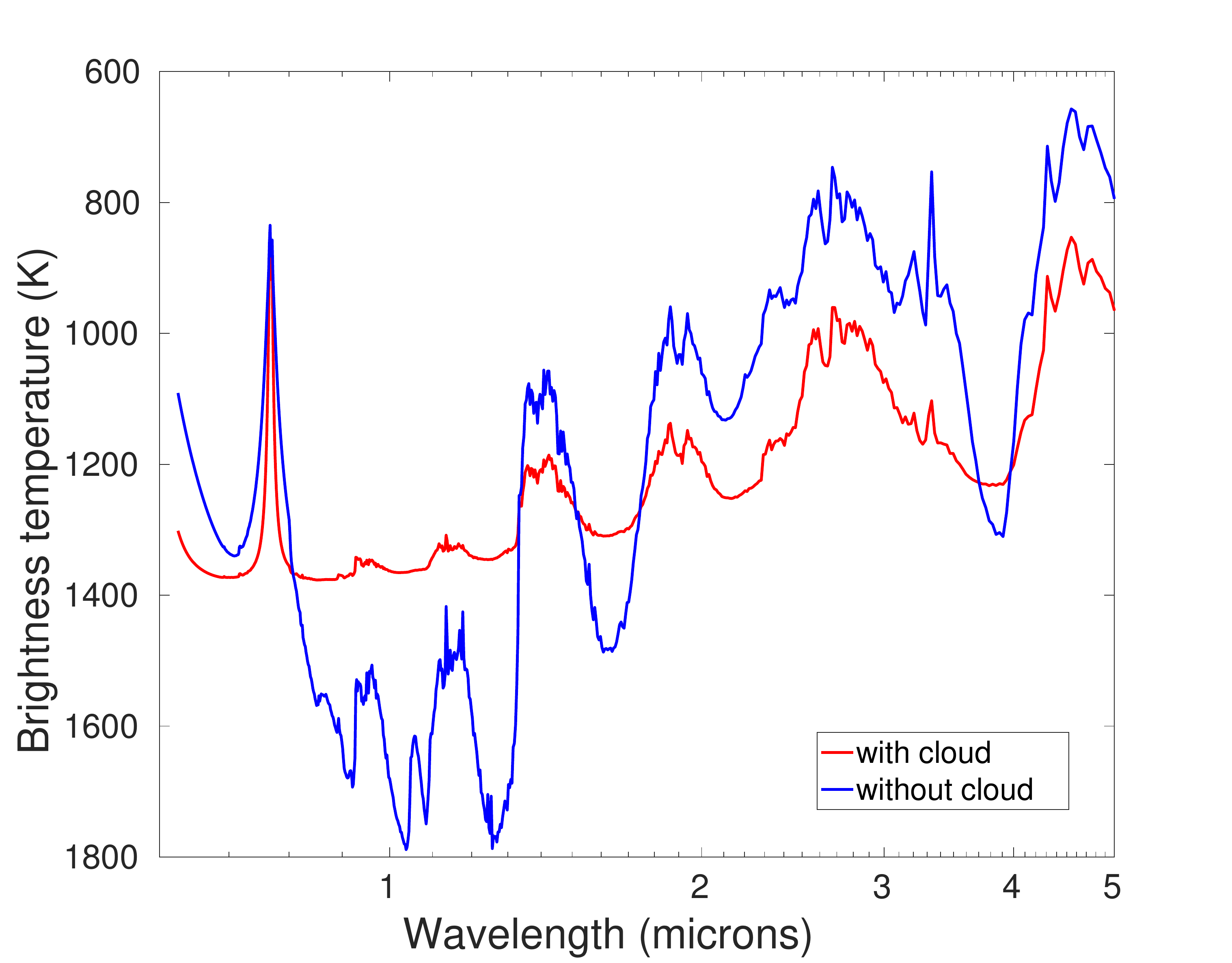}
\caption{Wavelength vs. brightness temperature for our best-fit ExoREM cloudy model (red curve) and an equivalent clear model (blue curve) with other parameters unchanged, showing where in the spectrum the brightness temperature increases/decreases with clouds.  In the cloudy case, the brightness temperature increases at longer 
wavelengths (e.g. 3 and 4.5 $\mu$m) by around 200 K relative to the clear case. The opposite is true at shorter wavelengths ($\sim$1-2 $\mu$m), where the brightness temperature decreases by 200 K relative to the clear case.  While we do not expect any fully clear patches on this object, longitudinal variations in the cloud thickness should produce similar trends and thus a $\sim$180$^{\circ}$ phase shift between near- and mid-IR lightcurves.}
\label{fig:tbright}       % Give a unique label
\end{figure*}

Longitudinal variations in cloud thickness can potentially produce anti-correlated variability, in other words $\sim$180$^{\circ}$ phase shifts between the near-IR and mid-IR lightcurves.  
In Fig.~\ref{fig:tbright}, we plot wavelength vs. brightness temperature, showing where in the spectrum the brightness temperature increases/decreases with clouds.  In the cloudy case, the brightness temperature increases at longer wavelengths (e.g. 3 and 4.5 $\mu$m) by around 200 K relative to the clear case. 
The opposite is true at shorter wavelengths ($\sim$1-2 $\mu$m), where the brightness temperature decreases by 200 K relative to the clear case.  
A hole in the cloud cover would thus produce a 180 phase shift between near-IR and mid-IR lightcurves, except for the 1.4~$\mu$m water band which would be correlated with the mid-IR lightcurve, contrary to the observations presented herein for PSO J318.5-22.
However, we do not expect any fully clear patches on this object \citep[and~indeed~previous~work~suggests~this~is~the~case~for~brown~dwarfs~in~general,][]{Apa13}, but rather longitudinal variations in the cloud thickness.  The phase shift and the amplitude of light curves are then very dependent on the altitude, thickness, and placement of different cloud species. 

\begin{figure*}
% For example, with the graphicx style use
%\includegraphics[scale=0.4]{p_photosphere_1150K_g33b.eps}
\includegraphics[width=\linewidth]{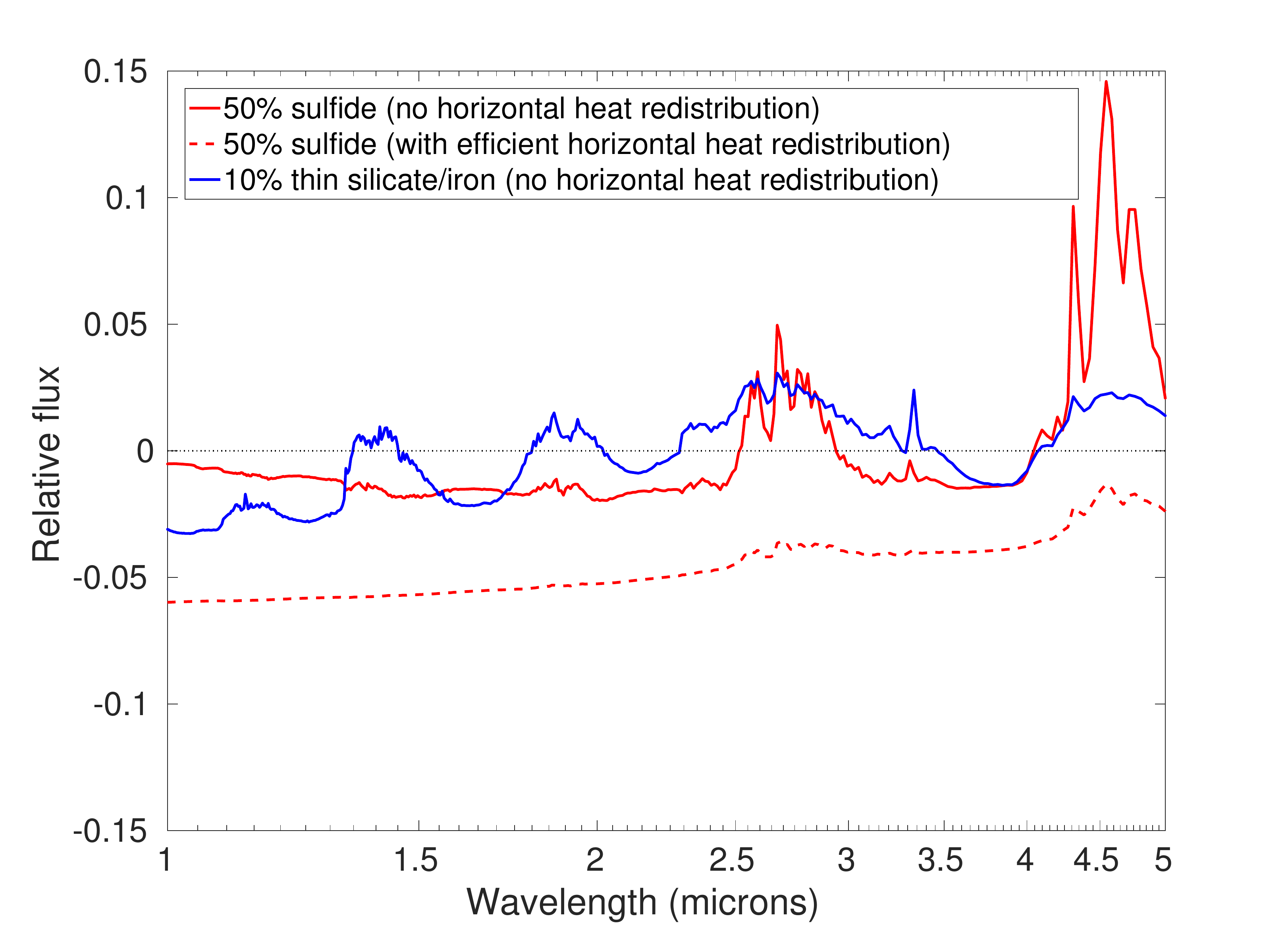}
\caption{Predicted light curve amplitude that would be produced assuming:  (1) a spot with optically thinner silicate and iron cloud thickness, covering 10\% of the surface, and homogeneous thick silicate / iron clouds on the remaining 90$\%$ (blue curve) and (2) one hemisphere covered by sulfide clouds and no sulfide clouds on the other hemisphere, with homogeneous silicate/iron clouds for both hemispheres (red curves).
Case (2a) was computed assuming no horizontal heat redistribution between the less cloudy spot and the rest of the brown dwarf. Case (2b) was computed with no horizontal heat redistribution (solid line) and with very efficient heat redistribution (dashed line).}
\label{fig:cloudflux}       % Give a unique label
\end{figure*}

%should produce similar trends and thus a $\sim$180$^{\circ}$ phase shift between near and mid-IR lightcurves.
We consider a number of simple geometries for both silicate / iron clouds and sulfide clouds to model our observed amplitudes and phases.
In Fig.~\ref{fig:cloudflux}, we computed the light curve amplitude that would be produced assuming:  (1) a spot with optically thinner silicate and iron cloud thickness, covering 10\% of the surface, with homogeneous thick silicate and iron clouds over the rest of the surface and 
(2) one hemisphere covered by high-altitude sulfide clouds and no sulfide clouds on the other hemisphere, with homogeneous silicate/iron clouds below the sulfide cloud layer altitude for both hemispheres.
Case (2a) was computed assuming no horizontal heat redistribution between the less cloudy spot and the rest of the brown dwarf. Case (2b) was computed with no horizontal heat redistribution (solid line) and with very efficient heat redistribution (dashed line).
For longitudinal variations (Case 1) in silicate and iron cloud thickness, we predict large variations in the amplitude within the 1.07 - 1.67 $\mu$m spectral range of the HST WFC3 G141 grism and, additionally, the 1.4~$\mu$m water band lightcurve should be correlated with the 4.5~$\mu$m Spitzer Channel 2 lightcurve.
For longitudinal variations in the sulfide cloud cover (Case 2a and b), the amplitude is quite constant in the HST bands and for a case intermediate between efficient heat redistribution and no heat redistribution, the predicted amplitude and the phase shifts between different near-IR and mid-IR wavelengths could be compatible with our HST and Spitzer observations.
This modelling remains very preliminary, but suggests that variations in the cover of high altitude clouds could begin to explain the observations presented herein. Na$_2$S clouds are a good candidate for such high altitude clouds since they form at very low pressures high in the atmosphere (0.06 bar). Inhomogeneous Cr and MnS clouds also are potential candidates. An upper layer of silicate clouds could also produce such variability but it would require a mechanism for forming or transporting cloud particle higher than the cloud deck. Cloud convection triggered by latent heat release \citep{Tan17} or radiative heating \citep{Fre10} may produce vertically extended clouds and a detached silicate haze layer.  

Our measured phase shifts between the HST bands and the Spitzer 4.5 $\mu$m lightcurve are in fact somewhat more than 180$^{\circ}$, which is unsurprising, as the cloud geometry for PSO J318.5-22 is certainly more complicated than the simple geometries considered above.    
%The further observed phase shift may also perhaps be a result of beat patterns produced by planetary-scale wave pairs in the mid-IR top-of-atmosphere layer that are moving with slightly different velocities due to zonal wind variations \citep{Apai17}.  
Modeling approaches that combine multiple 1-D models \citep[such~as~the~one~presented~herein~and][]{Art09,Apa13,Kar15,Kar16} can reproduce correlated variability or 180$^{\circ}$ anti-correlated variability, but not other phase shifts.   
%Approaches for modeling the surface features responsible for the observed variability assuming a single "top-of-atmosphere" layer that can be modeled using a combination of 1-D models \citep{Apa13,Kar15,Kar16}.
As demonstrated above, where the "top-of-atmosphere" occurs varies depending on wavelength and the specific opacity sources that dominate at different atmospheric levels.  The observed "phase shifts" may simply be heterogeneous and uncorrelated structure at different altitudes, which is still modulated by the rotation period of the object in question.
%Thus, the measured phase shifts make fitting combinations of 1-d models as in \citet{Apa13} problematic -- such analyses assume one "top-of-atmosphere" pressure level and similar behavior at all altitudes.  Such combinations of 1-d models can reproduce correlated variability or 180$^{\circ}$ anti-correlated variability, but not other phase shifts.  Clearly 
Likely full 3-d models will be necessary to describe this structure, such as the C05BOLD model currently undergoing testing (Allard et al. in prep), especially as rotation probably plays a significant role in the appearance and features of these atmospheres \citep{Sho13}.

%\citet{Tre15} and \citet{Tre16} cloud-free models -- can these produce the observed variability?

\subsection{Variability in Low-Surface Gravity L dwarfs}

\citet{Met15} find increased mid-IR variability amplitudes for 8 low-surface gravity L3 to L5 objects with respect to the rest of their older, high-surface gravity survey sample.  We tentatively find that such a trend (in both the near and mid-IR) may continue for low-surface gravity mid-to-late L dwarfs.  Only 4 such objects have been surveyed in either the near or mid-IR to date.  Three out of the 4 have positive variability detections in both near- and mid-IR \citep[][]{Bil15,Lew16,Vos17b,Mor06}; one is a non-detection in our ongoing SofI survey (Vos et al. in prep).  
The three low-surface gravity mid-to-late L dwarfs with positive variability detections are PSO J318.5-22, W0047, and 2M2244.  For our HST+Spitzer monitoring of PSO J318.5-22, we found peak-to-trough amplitudes of $\sim$3.4$\%$ for Spitzer Channel 2 and 4.4 - 5.8$\%$ in the near-IR band (1.07-1.67 $\mu$m) covered by the WFC3 G141 prism.  In the discovery epoch, \citet{Bil15} found peak-to-trough variability amplitudes of 7-10$\%$ in the $J_S$ band and $\sim$3$\%$ in $K_S$, indicating evolution of the variability between the discovery epoch and our HST observations.  The lower amplitude in $K$ vs. $J$ during the discovery epoch is consistent with our finding in this work that variability amplitude decreases with increasing wavelength across the 1.1-1.7 $\mu$m spectral range of the HST WFC3 grism.  \citet{Lew16} find a similarly high near-IR variability amplitude for W0047, with the relative variability amplitude decreasing from 11$\%$ at 1.1 $\mu$m to 6.5$\%$ at 1.7 $\mu$m. 
\citet{Vos17b} reported a mid-IR detection for this object with a relative variability amplitude of 1.07$\pm$0.04$\%$.  \citet{Mor06} measured a Spitzer channel 1 peak-to-peak variability amplitude of 8 mmag for 2M2244, an L6.5 AB Dor member \citep{Vos17b}.  Variability in this object has recently been confirmed by \citep{Vos17b} who found a Spitzer channel 1 peak-to-peak variability amplitude of 0.8$\pm$0.2$\%$ and $\geq$3$\%$ variability amplitude in J band in a 4-hour long, $J$ band UKIRT WFCAM observation of this object.  All three of these objects have notably high near-IR amplitudes compared to field brown dwarfs with similar spectral types as well as planetary mass objects with earlier spectral types.  For instance, the detection of variability in the L5 planetary mass object 2M1207b has a considerably lower near-IR amplitude of 1-2\% (Zhou et al. 2016).  Of the three, PSO J318.5 also has a notably high mid-IR amplitude; mid-IR amplitudes for the other two objects are more in line with typical values for field brown dwarfs.   

With such a small number of low-surface gravity mid-to-late L variables to study, 
it is not clear whether these three objects are unusual -- or if low-surface gravity objects are inherently more variable then their high surface gravity counterparts.  The peak-of-variability for field brown dwarfs appears to be at the L/T transition \citep{Rad14a,Rad14b}, commonly attributed to the breakup or at least thinning of silicate clouds at this spectral type transition \citet{Apa13}.  It is hard to say if this is the case for low surface gravity objects, with three high-amplitude near-IR detections for mid-to-late L low surface gravity objects \citep[][]{Bil15,Zho16,Vos17b}, one high-amplitude detection in a young T2.5 object \citep{Art09,Gagn17}, and one tentative detection in a young T3.5 object \citep{Naud17}.
Predominantly early L low-surface gravity objects have been surveyed to date (Vos et al. in prep), largely due to the current scarcity of late L, L/T transition, and T spectral type young, low-surface gravity objects. Nonetheless, the few mid-to-late L objects surveyed to date appear to be notably variable, which is surprising given that late-L objects are expected to have thick (and probably homogeneous) cloud cover. If late-L spectral type young objects are as a class highly variable, this may draw into question the interpretation of high amplitude variability as
the breakup of silicate clouds between the L and T spectral type.

Low-surface gravity mid-to-late L dwarfs are particularly interesting because these objects are excellent proxies for several known giant exoplanet companions.  The spectra of PSO J318.5-22 and W0047 are nearly identical to those of inner two HR 8799 planets \citep{Bon16}.  \citet{DeR16} find that the spectrum of the particularly red planet HIP 95086b \citep{Ram13} closely matches that of 2M2244.
The newly-discovered exoplanet companion HIP65426b also has an L5-L7 spectral type \citep{Chau17}.  Given the significant variability of PSO J318.5-22, W0047, and 2M2244, we may expect exoplanet companions such as HR 8799bcde, HIP 95086b, and HIP 65426b to be similarly variable, although viewing angle (likely pole-on for the HR 8799 system) may render that variability hard to detect.

\subsection{Are Young, Planetary Mass Objects Fast Rotators?}

\begin{figure*}
% For example, with the graphicx style use
%\includegraphics[scale=0.4]{p_photosphere_1150K_g33b.eps}
\includegraphics[width=\linewidth]{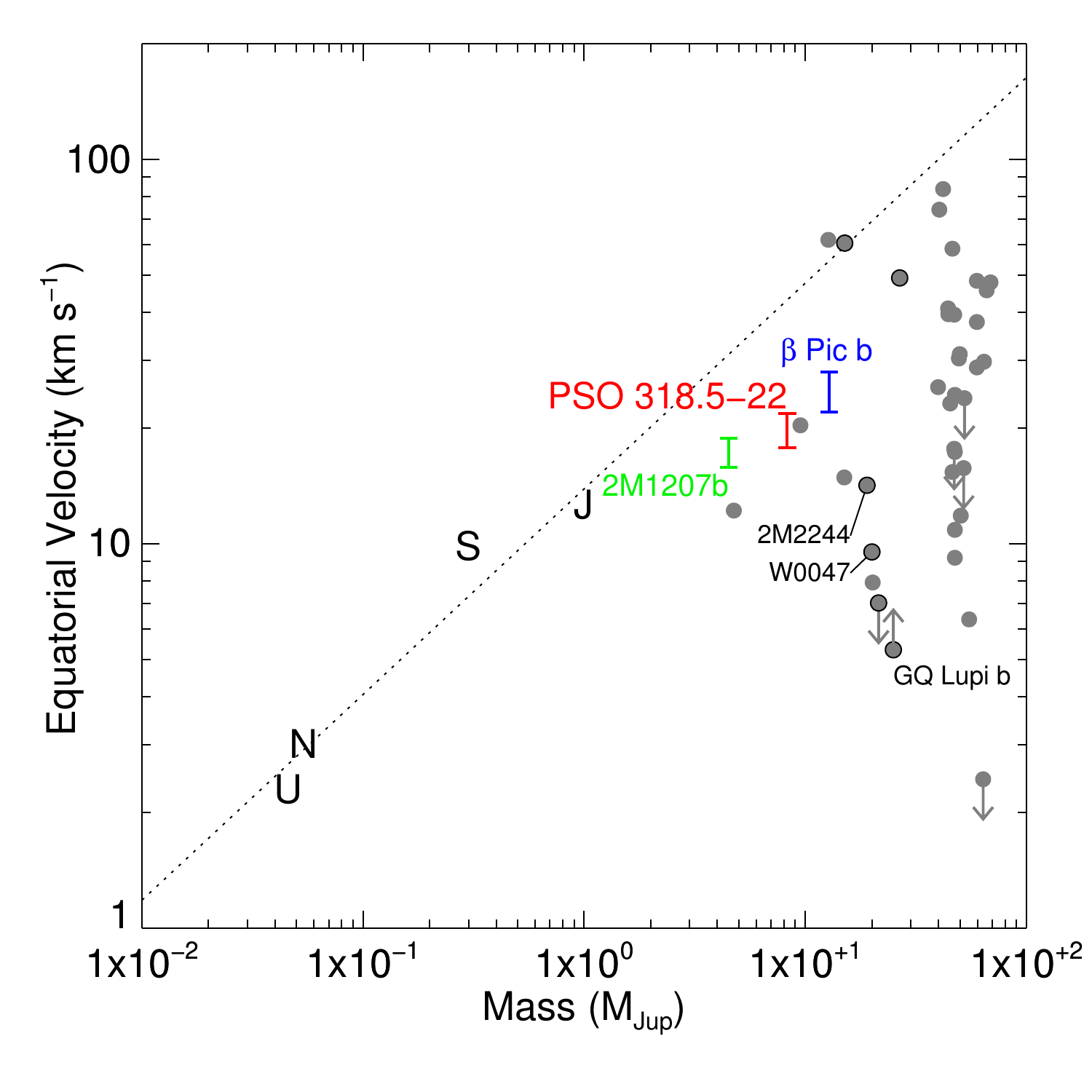}
\caption{Mass vs. equatorial velocity for young, planetary mass objects including PSO J318.5-22, as well as the exoplanet $\beta$ Pic b, 2M1207b, and two $\sim$20 M$_{Jup}$ members of AB Dor, 2M2244 and W0047.  Solar system planets and
brown dwarfs with measured periods from \citet{Vos17} are plotted as gray circles; young brown dwarfs are plotted as gray circles outlined in black.
Planetary mass objects seem to encompass a similar range of equatorial velocities as older, field brown dwarfs, with both rapid rotators and notable slow rotators such as the young, 30-40 M$_{Jup}$ brown dwarf companion GQ Lup b \citep{Sch16}. 
}
\label{fig:massvsvrot}       % Give a unique label
\end{figure*}

Even if young, planetary mass objects have significant top-of-atmosphere inhomogeneities, we will only be able to detect such features if these objects are relatively rapid rotators (periods $<$20 hours or so). Many old, field brown dwarfs are rapid rotators \citep{Zap06}.  From conservation of angular momentum, one might expect young objects to be predominantly slower rotators compared to old, field brown dwarfs, as they have somewhat inflated radii \citep[e.g.~$\sim$1.4~R$_{Jup}$~for~PSO J318.5-22][]{All16} compared to older objects (radius~$\sim$1 R$_{Jup}$) and will be expected to spin up with age as they contract.  At least preliminarily, however, there is a small cohort of young ($\leq$150 Myr), planetary mass objects with periods $<$20 hours, including PSO J318.5-22, as well as the bonafide exoplanet $\beta$ Pic b and 2M1207b.  In Fig.~\ref{fig:massvsvrot}, we plot estimated object mass vs. measured equatorial velocity for these objects, solar system objects, and field brown dwarfs with measured periods from \citet{Vos17}.  Planetary mass objects seem to encompass a similar range of equatorial velocities as older, field brown dwarfs, with both rapid rotators and notable slow rotators such as the young, 30-40 M$_{Jup}$ brown dwarf companion GQ Lup b \citep{Sch16}.  However, statistics are still too sparse for a robust comparison to the brown dwarf population in general.  Preliminary analyses do suggest that the rotation rate between free-floating and companion objects is similar -- \citet{Brya17} recently measured rotation rate for a number of companions with masses $<$20 M$_{Jup}$.  Combining their measurements with others in the literature, they found no discernable difference in rotation speed between companions and free-floating objects with similar masses for a small sample of 11 objects.

%Vos et al. resubmitted measure a 3.6 $\mu$m Spitzer variability amplitude of $\sim$1$\%$ for the young planetary mass analogue W0047.  

\section{Conclusions}

Here we present simultaneous HST WFC3 + Spitzer IRAC variability monitoring 
for the variable planetary mass object PSO J318.5-22.  Our simultaneous HST + Spitzer observations covered $>$2 rotation periods with Spitzer and most of a rotation period with HST.  The main results from these observations are: 

\begin{itemize}
\item Detection of high amplitude variability in both near-IR and mid-IR bands with a period of 8.6$\pm$0.1 hours.  We estimate peak-to-trough variability amplitudes of 3.4$\pm$0.1$\%$ for Spitzer Channel 2 and 4.4 - 5.8$\%$ (typical uncertainty of $\sim$0.3$\%$) in the near-IR bands (1.07-1.67 $\mu$m) covered by the WFC3 G141 prism.
\item a relatively high inclination for PSO J318.5-22 of 56$\pm$8$^{\circ}$, derived by
combining our measured period with the measured $v sin i$ from \citep{All16} for this object.
Thus, we are observing close to the full intrinsic variability amplitude in each band.    
\item Detection of 200-210$^{\circ}$ (typical uncertainty of $\sim$4$\%$) phase offsets between the near-IR and mid-IR lightcurves, likely indicating varying longitudinal atmospheric structure at different depths in this atmosphere
\item Tentative detection of a small $\sim$6$^{\circ}$ phase offset between the 2MASS J band and the rest of the near-IR bands, but this is at a considerably lower significance level than the mid-IR vs. near-IR phase shift
\item A decrease of variability amplitude as a function of increasing wavelength, as has previously been found for field brown dwarfs \citep[c.f.~among~others][]{Apa13,Rad14a,Yan16}. 
We tentatively find that the amplitude of variability in the 1.4 $\mu$m water absorption feature is slightly smaller than adjacent wavelengths in the first orbit of our observations, but similar to adjacent wavelengths in the final orbit of our observations.
\item Detection of similar variability amplitudes in wide spectral bands relative to absorption features, suggesting that the driver of the variability may be inhomogeneous clouds (perhaps variations in the cover of high altitude clouds over a homogeneous layer of thick clouds) as opposed to hot spots or compositional inhomogeneities at the top-of-atmosphere level.  Na$_2$S clouds are a good candidate high altitude cloud species since they form at very low pressures high in the atmosphere (0.06 bar). Inhomogeneous Cr and MnS clouds also are potential candidates. 
\end{itemize}

%The wavelength dependence of the observed variability resembles that of mid-L dwarfs studied by \citet{Yan15}, with amplitude decreasing monotonically with increasing wavelength, rather than that found for L/T transition brown dwarfs, which show smaller variability amplitudes at 1.4$\mu$m relative to the rest of the near-IR \citet{Apa13}.  
Both mid-IR and near-IR variability amplitudes for PSO J318.5-22 are large -- comparable with that of high-amplitude L/T transition brown dwarfs and considerably larger than found for the early or mid-L dwarfs \citep{Yan15,Rad14a,Met15}.  Clearly, while low surface gravity late-L planetary analogues share some variability properties with field brown dwarfs, they are their own unique category of objects and merit the same in-depth observation and analysis.  Given the significant variability of PSO J318.5-22 and other mid-to-late L low surface gravity objects, we may expect variability as well in exoplanet companions such as HR 8799bcde, HIP 95086b, and HIP 65426b which share similar spectral types and surface gravities.

\acknowledgments
Based on observations made with the NASA/ESA Hubble Space Telescope, obtained at the Space Telescope Science Institute, which is operated by the Association of Universities for Research in Astronomy, Inc., under NASA contract NAS 5-26555. These observations are associated with program \# 14188.  KNA acknowledges support for program \#14188 provided by NASA through a grant from the Space Telescope Science Institute, which is operated by the Association of Universities for Research in Astronomy, Inc., under NASA contract NAS 5-26555.  This work is based in part on observations made with the Spitzer Space Telescope, which is operated by the Jet Propulsion Laboratory, California Institute of Technology under a contract with NASA.  BAB and JV also acknowledge support from STFC grant ST/J001422/1.  We thank Jack Gallimore for providing the posterior $v sin i$ distribution for PSO J318.5-22 and Mike Cushing for a close reading of this manuscript and useful conversations.

\bibliographystyle{aasjournal}
\bibliography{main} % if your bibtex file is called example.bib

\clearpage

\clearpage

\clearpage

\clearpage

\clearpage

\clearpage

\clearpage

\clearpage

\clearpage

\clearpage

\clearpage

\clearpage

\clearpage

\clearpage

\clearpage

\appendix

\section{HST lightcurve MCMC posteriors \label{sec:appendix}}

\begin{figure*}
% For example, with the graphicx style use
\includegraphics[scale=0.7]{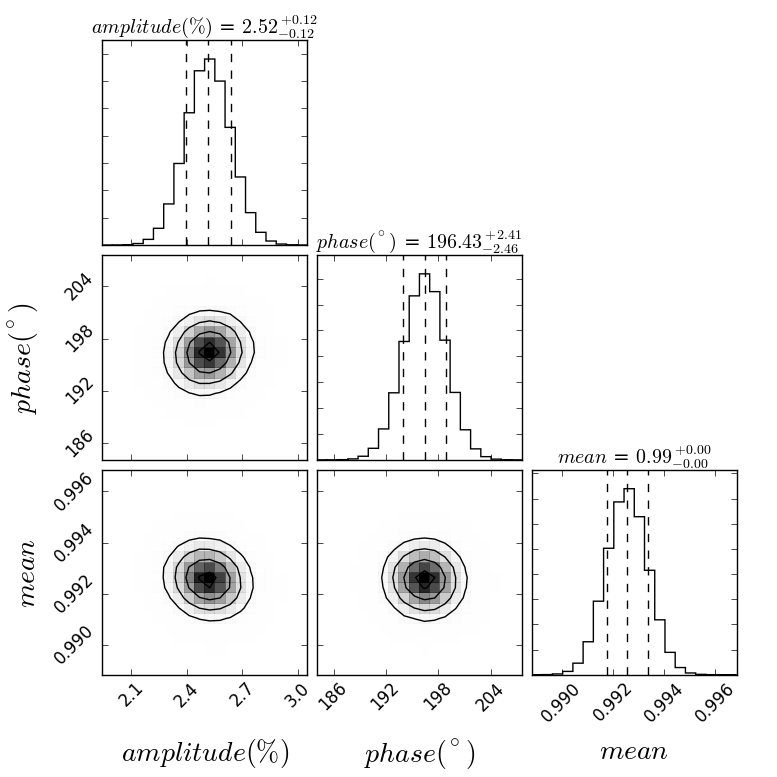}
\caption{Posterior probability distributions of parameters from sinusoid MCMC fits to the HST "white light" lightcurve (full bandpass from 1.07 to 1.67 $\mu$m) for PSO J318.5-22.  Since the HST observation does not cover a full rotation period, we have fixed the period to 8.6 hours, as found from the Spitzer lightcurve. In the marginalized confidence interval plots, the middle dashed line gives the median, the two outer vertical dashed lines represent the 68\% confidence interval. The contours show the 1, 1.5 and 2-$\sigma$ levels.
}
\label{fig:HST_wllc_mcmc}       % Give a unique label
\end{figure*}

\begin{figure*}
% For example, with the graphicx style use
\includegraphics[scale=0.7]{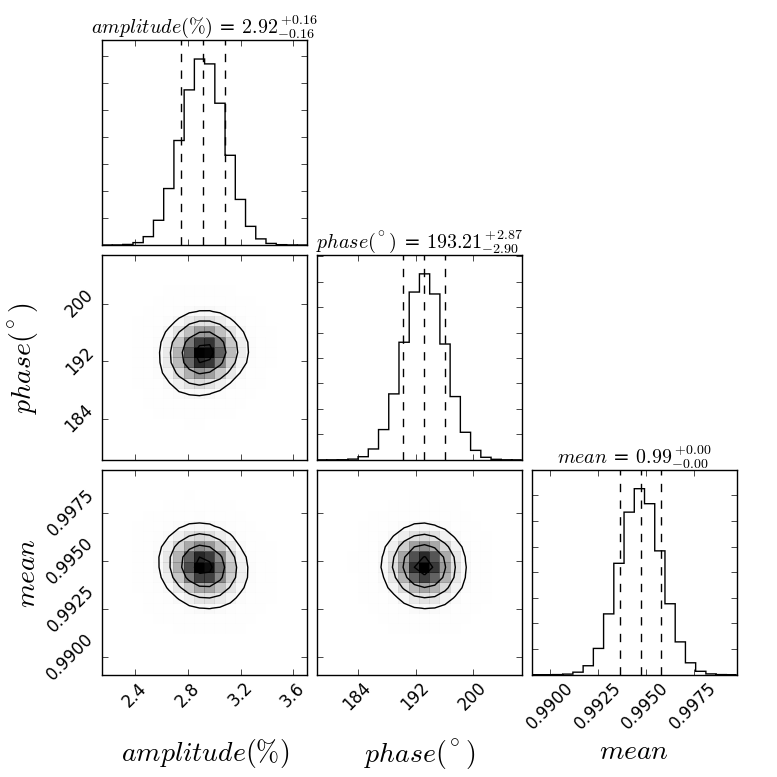}
\caption{Posterior probability distributions of parameters from sinusoid MCMC fits to the HST synthesized 2MASS J lightcurve for PSO J318.5-22.  Since the HST observation does not cover a full rotation period, we have fixed the period to 8.6 hours, as found from the Spitzer lightcurve. In the marginalized confidence interval plots, the middle dashed line gives the median, the two outer vertical dashed lines represent the 68\% confidence interval. The contours show the 1, 1.5 and 2-$\sigma$ levels.
}
\label{fig:HST_J_mcmc}       % Give a unique label
\end{figure*}

\begin{figure*}
% For example, with the graphicx style use
\includegraphics[scale=0.7]{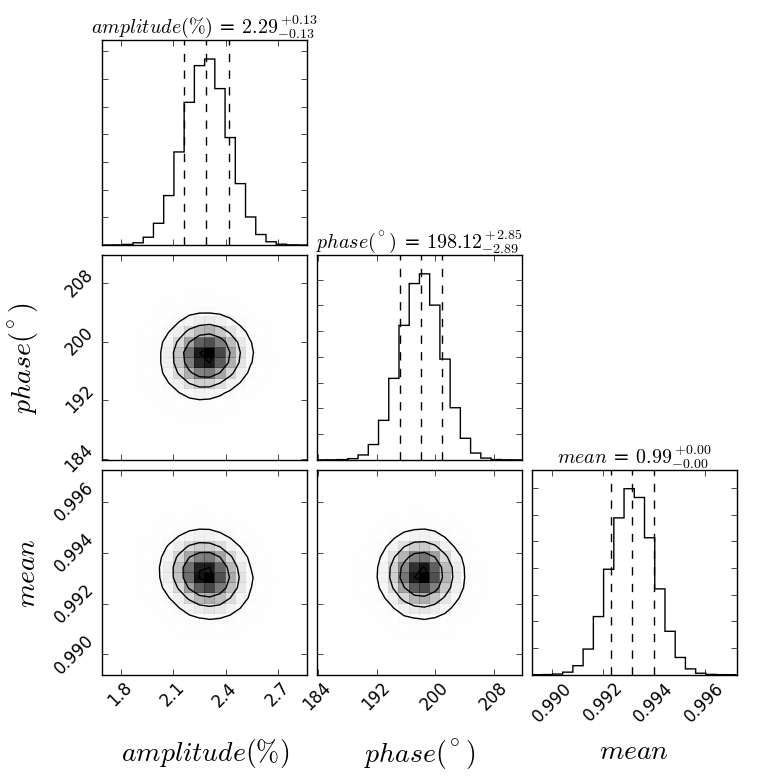}
\caption{Same as Fig.~\ref{fig:HST_J_mcmc}, for the HST synthesized 2MASS H lightcurve.
}
\label{fig:HST_H_mcmc}       % Give a unique label
\end{figure*}

\begin{figure*}
% For example, with the graphicx style use
\includegraphics[scale=0.7]{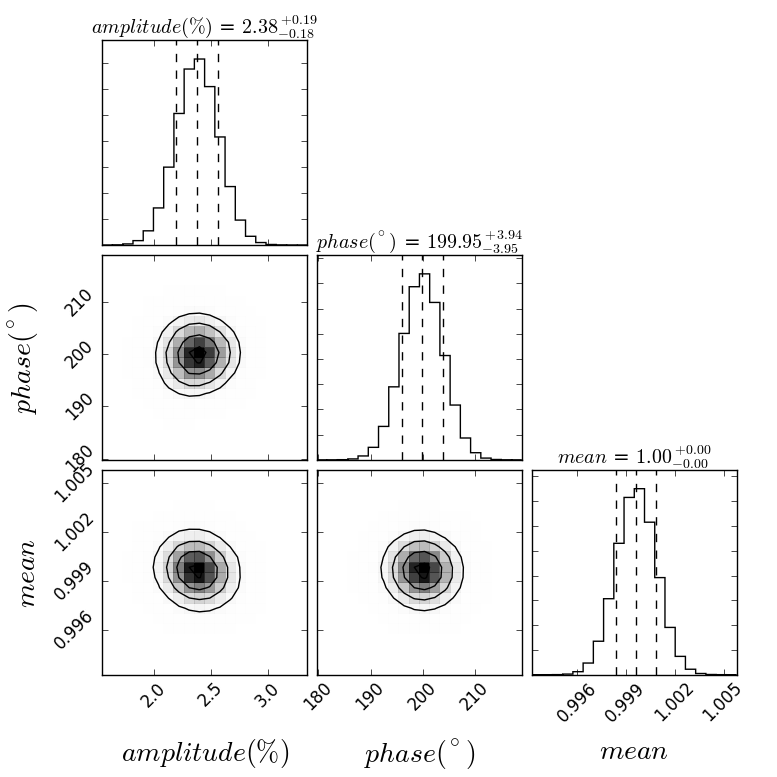}
\caption{Same as Fig.~\ref{fig:HST_J_mcmc}, for the HST synthesized water band lightcurve (1.34 to 1.44 $\mu$m). 
}
\label{fig:HST_water_mcmc}       % Give a unique label
\end{figure*}

\begin{figure*}
% For example, with the graphicx style use
\includegraphics[scale=0.7]{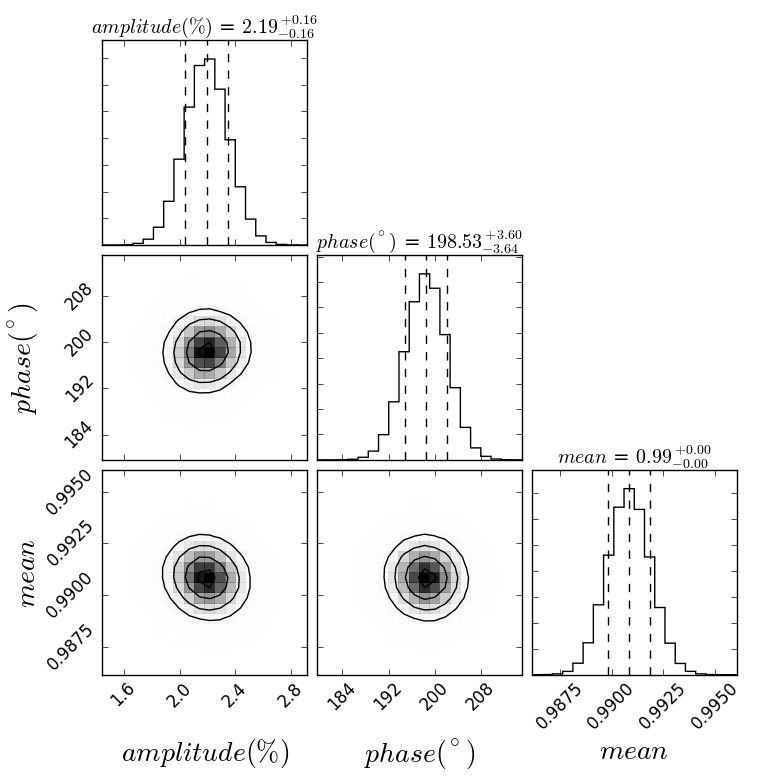}
\caption{Same as Fig.~\ref{fig:HST_J_mcmc}, for the HST synthesized methane band lightcurve (1.60 to 1.67 $\mu$m).
}
\label{fig:HST_methane_mcmc}       % Give a unique label
\end{figure*}

MCMC posteriors for sinusoidal fits to HST lightcurves are presented in Fig.~\ref{fig:HST_wllc_mcmc} through Fig.~\ref{fig:HST_methane_mcmc}.

\clearpage

\vspace{5mm}
\facilities{HST(WFC3), Spitzer(IRAC)}
\software{python, astropy, IDL, emcee}


%\software{astropy \citep{2013A&A...558A..33A},  
%          Cloudy \citep{2013RMxAA..49..137F}, 
%          SExtractor \citep{1996A&AS..117..393B}
%          }





\end{document}